\documentclass[aps,pra,twocolumn,amsmath,amssymb,bibliography,superscriptaddress]{revtex4-2}
\usepackage{times}
\usepackage[english]{babel}
\usepackage{latexsym}
\usepackage{graphics}
\usepackage{subfigure}
\usepackage{epsfig}
\usepackage{color}
\usepackage{hyperref}
\usepackage{braket} 
\usepackage{amstext}
\usepackage{amssymb}
\usepackage{graphicx}
\usepackage{esint}
\usepackage{braket}
\usepackage{babel}
\usepackage{amsmath}
\usepackage{float}
\hypersetup{
colorlinks=true,
citecolor=blue,
linkcolor=blue,
urlcolor=blue
}

\makeatletter
\renewcommand{\fnum@figure}{FIG.~\thefigure}
\renewcommand{\fnum@table}{TABLE.~\thetable}
\makeatother

\usepackage{todonotes}

\begin{document}
\title{
Overcoming disorder in superconducting globally-driven quantum computing
}

\author{Riccardo Aiudi}
\affiliation{Planckian srl, I-56127 Pisa, Italy}
\author{Julien Despres}
\affiliation{Planckian srl, I-56127 Pisa, Italy}
\author{Roberto Menta}
\email{rmenta@planckian.co}
\affiliation{Planckian srl, I-56127 Pisa, Italy}
\affiliation{NEST, Scuola Normale Superiore, I-56127 Pisa, Italy}
\author{Ashkan Abedi}
\affiliation{Planckian srl, I-56127 Pisa, Italy}
\affiliation{NEST, Scuola Normale Superiore, I-56127 Pisa, Italy}
\author{Guido Menichetti}
\affiliation{Planckian srl, I-56127 Pisa, Italy}
\affiliation{Dipartimento di Fisica dell’Universit\`{a} di Pisa, Largo Bruno Pontecorvo 3, I-56127 Pisa, Italy}
\author{Vittorio Giovannetti}
\affiliation{Planckian srl, I-56127 Pisa, Italy}
\affiliation{NEST, Scuola Normale Superiore, I-56127 Pisa, Italy}
\author{Marco Polini}
\affiliation{Planckian srl, I-56127 Pisa, Italy}
\affiliation{Dipartimento di Fisica dell’Universit\`{a} di Pisa, Largo Bruno Pontecorvo 3, I-56127 Pisa, Italy}
\author{Francesco Caravelli}
\email{fcaravelli@planckian.co}
\affiliation{Planckian srl, I-56127 Pisa, Italy}
\affiliation{Dipartimento di Fisica dell’Universit\`{a} di Pisa, Largo Bruno Pontecorvo 3, I-56127 Pisa, Italy}

\begin{abstract} 
We study the impact of static disorder on a globally-controlled superconducting quantum computing architecture based on a quasi-two-dimensional ladder geometry [\href{https://doi.org/10.1103/PhysRevResearch.7.L012065}{Phys. Rev. Research~{\bf 7}, L012065 (2025)}]. Specifically, we examine how fabrication-induced inhomogeneities in qubit resonant frequencies and coupling strengths affect quantum state propagation and the fidelity of fundamental quantum operations. Using numerical simulations, we quantify the degradation in performance due to disorder and identify single-qubit rotations, two-qubit entangling gates, and quantum information transport as particularly susceptible. To address this challenge, we rely on pulse optimization schemes, and, in particular, on the \texttt{GRAPE} (Gradient Ascent Pulse Engineering) algorithm. Our results demonstrate that, even for realistic levels of disorder, optimized pulse sequences can achieve high-fidelity operations, exceeding 99.9\% for the three quantum operations, restoring reliable universal quantum logic and robust information flow. These findings highlight pulse optimization as a powerful strategy to enhance the resilience to disorder of solid-state globally-driven quantum computing platforms.
\end{abstract}

\maketitle

\section{Introduction}

The rapid progress of quantum computing demands scalable architectures capable of high-fidelity control over large qubit arrays. Superconducting circuits, among quantum computing platforms, can offer both potential scalability and compatibility with established fabrication processes~\cite{Martinis2019quantum, supremacy_PRL_2021, bravyi2022future, ezratty2023perspective}. However, conventional methods are based on individually addressable control lines, and while offering great flexibility, this leads to significant wiring overhead and crosstalk~\cite{Girvin2008, Gambetta2017}. As processor sizes increase, these issues intensify, highlighting the need for more efficient and robust qubit control schemes~\cite{Kjaergaard2020, Mohseni2024}.

A promising direction is the adoption of globally-controlled \textit{digital} architectures, where a shared control field simultaneously addresses multiple qubits, minimizing hardware complexity and facilitating scalability. This concept was first introduced by Lloyd in 1993~\cite{Lloyd_1993, Lloyd_1993_SI}, and later developed extensively by Benjamin and collaborators~\cite{benjamin_2001_1, benjamin_2001_2, Levy_2002, benjamin_2003, benjamin_2004, benjamin-bose_2004, Ivanyos_2005, Kay_2004, Fitzsimons_2006, Silva_2009}. The idea of encoding quantum degrees of freedom has also been considered in the literature \cite{viola}. Despite their conceptual appeal, these schemes remained largely theoretical until recently. Experimental exploration and suggestions of this type of architectures has now begun with various platforms, including Rydberg atoms~\cite{cesa2023universal}, spin qubits~\cite{Patomaki_2024}, and superconducting circuits~\cite{menta2024globally, cioni2024conveyorbelt, menta2025building}.

While global control reduces the need for independent control lines, its resilience to imperfections, such as static disorder in superconducting qubits, remains a major challenge. This stems from the requirement for distinct qubit species and uniform, always-on nearest-neighbor interactions~\cite{menta2025building}. Such uniformity is difficult to achieve experimentally, as fabrication inevitably introduces variations in qubit resonant frequencies, coupling strengths, and crosstalks.

In the present Article, we investigate the impact of disorder on a globally-driven superconducting quantum computing architecture based on a 2D ladder geometry~\cite{menta2024globally} and demonstrate how its detrimental effect can be effectively mitigated. The ladder architecture harnesses the effects of ZZ interactions and an emergent blockade regime. The same investigation is also easily applicable to the conveyor-belt architecture discussed in Ref.~\cite{cioni2024conveyorbelt}or even to different geometries~\cite{menta2025building}. Although globally-controlled architectures offer a promising path towards scalable quantum computing, their practical implementation must address key challenges. In both ladder~\cite{menta2024globally} and conveyor-belt~\cite{cioni2024conveyorbelt} architectures, static disorder arising from fabrication-induced variations in qubit parameters can disrupt coherent quantum information transfer and degrade the fidelity of quantum operations.

To address this issue, we systematically analyze how disorder influences quantum state propagation along a two-row ladder and assess its effects on the performance of single- and two-qubit gate operations. Furthermore, we explore pulse optimization techniques as a strategy to counteract disorder-induced errors, demonstrating that, despite inherent system imperfections, good fidelities can still be achieved. Our results support the viability of globally-controlled architectures by showing that disorder can be mitigated through pulse optimization, supporting the broader effort of designing robust and scalable quantum processors. 

The remainder of this Article is structured as follows: Sec.~\ref{sec:intro-global-main} introduces the theoretical framework and describes the globally controlled architecture under consideration; Sec.~\ref{sec:disordermain} presents numerical simulations and analyzes disorder effects; while Sec.~\ref{sec:pulse-optimization-main} discusses the implementation of pulse optimization strategies to overcome disorder (the main result of this work). We end with Sec.~\ref{sec:conclusion} summarizing our results and outlining potential future research directions.

\vspace{1.5cm}

\section{Ladder quantum architecture}\label{sec:intro-global-main}

\begin{figure}[t]
\centering
\includegraphics[width=0.98\columnwidth]{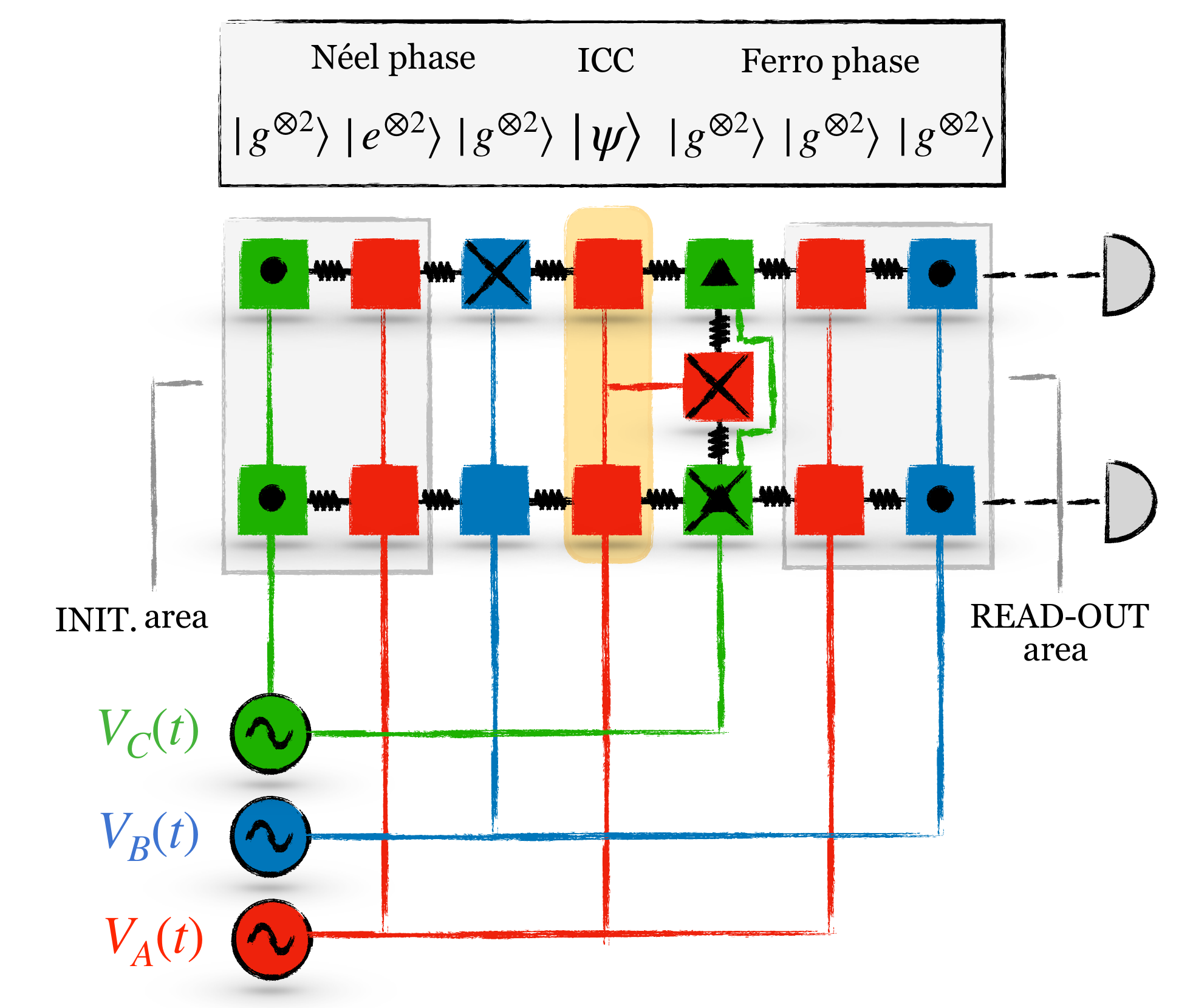}
\caption{\label{fig:ladder}~\textbf{Illustration of a 2D ladder globally driven quantum computing architecture.} Three types of superconducting qubits, $A$, $B$, and $C$ (red, blue, and green squares), are driven by separate classical sources $V_{A,B,C}(t)$ (colored lines) and coupled via longitudinal ZZ interactions (black springs). Crossed squares indicate qubits with double Rabi frequency~\cite{menta2024globally, menta2025building}. Black triangles and circles denote local resonance frequency corrections. $B$- and $C$-type intra-row crossed qubits enable single-qubit gates, while the $A$-type inter-row crossed qubit enables two-qubit gates. The yellow-highlighted column marks the ``information carrier column'' (ICC), interfacing N\'eel ($|geg\rangle$) and Ferro ($|ggg\rangle$) phases. The first two columns form the initialization (INIT.) area and the last two form the read-out area.
The figure corresponds to $N=2$ computational qubits quantum processor, and $15$ physical qubits.}
\end{figure}

We consider a superconducting quantum architecture based on a globally-driven ladder geometry introduced in Ref.~\cite{menta2024globally}, and shown in Fig.~\ref{fig:ladder}. The system comprises a 2D lattice of superconducting qubits arranged in $N$ horizontal rows and $2N{+}3$ vertical columns. Qubits are divided into three species, $A$, $B$, and $C$, each characterized by a distinct internal frequency $\omega_\chi$ and driven by a global time-dependent signal $V_\chi(t)$ with Rabi frequency $\Omega_\chi(t)$. Qubits of type $\chi$ are subject to the same control, and their arrangement follows a periodic intra-row pattern $CABACABA\ldots$. The total number of physical qubits, including $N-1$ mediating $A$-type qubits between neighboring rows, is $N_{\rm tot} = 2N^2 + 4N - 1$, scaling as $\mathcal{O}(N^2)$. Further architectural details are presented in App.~\ref{sec:app-global}. It is important to stress that while the number of qubits in this architecture scales quadratically in the number of computational qubits, there are alternative globally controlled designs that offer a linear scaling~\cite{cioni2024conveyorbelt, menta2025building}.

The Hamiltonian of the system is based on ZZ-type of interactions, and can be written in the form $\hat{H}(t) = \hat{H}_0 + \hat{H}_{\rm drive}(t)$, where 
\begin{align}
\label{eq:H0main}
    \hat{H}_0 &= \sum_{\chi \in \mathcal{S}} \sum_{i \in \chi} \frac{\hbar \omega_i}{2} \hat{\sigma}^{(z)}_i + \sum_{\langle i,j\rangle} \frac{\hbar \zeta}{2} \hat{\sigma}^{(z)}_i \hat{\sigma}^{(z)}_j, \\
\label{eq:HdriveMain}
    \hat{H}_{\rm drive}(t) &= \sum_{\chi \in \mathcal{S}} \sum_{i \in \chi} \hbar \Omega_\chi(t) \sin(\omega_{{\rm d},\chi} t + \phi_\chi(t)) \hat{\sigma}^{(y)}_i.
\end{align}
Here, $\mathcal{S} = \{A,B,C\}$, $\hat{\sigma}^{(\alpha)}_i$ refers to the Pauli operators, $\zeta$ denotes the ZZ coupling strength, and $\omega_i$ includes local detunings due to connectivity. The control fields $\Omega_\chi(t)$ and $\phi_\chi(t)$ are independent of qubit position, reflecting the global nature of the control. Crossed qubits (denoted $\chi^\times$) have a doubled coupling to the control line with respect to regular (non-crossed) qubits, $\chi^{\rm r}$, i.e. $\Omega_{\chi^\times} = 2\Omega_{\chi^{\rm r}}$, and enable localized gate operations within the globally driven scheme~\cite{menta2025building}. As shown in Ref.~\cite{menta2024globally}, the Hamiltonian above can be expressed in the rotating frame (RF), after performing the rotating-wave approximation (RWA), placing each qubit in a frame rotating at each species' drive frequency. After the latter, the hamiltonian reads:
\begin{align}
& \hat{H}_{\text{rf}}(t) \simeq \sum_{\chi \in \mathcal{S}} \sum_{i \in \chi} \frac{\hbar \Omega_\chi(t)}{2} \Big[e^{i\phi_{\chi}(t)}\vert g_i \rangle \langle e_i \vert + {\rm h.c.} \Big] \nonumber \\
& ~~~~~~~~~~~~~~+ \sum_{\langle i,j \rangle} 2 \hbar \zeta \vert e_i e_{j} \rangle \langle e_i e_{j} \vert,
\label{eq:RWARFhammain}
\end{align}  
in which one can immediately see the blockade effect, mimicking the one of the Rydberg atoms, and being crucial for the quantum operations of the ladder architecture.
See App.~\ref{app:model} for the full model and derivations.  

From the point of view of the computing architecture, each row encodes a computational qubit by localizing quantum information at the boundary between a ferromagnetic domain ($|ggg\rangle$) and a N\'eel-ordered domain ($|geg\rangle$), forming an \textit{information-carrying column} (ICC). The ICC can be coherently shifted along the row by a sequence of global pulses, effectively moving the computational qubit while preserving its quantum state (see App.~\ref{app:infoflow} for details). This dynamical information transport relies on the interaction blockade regime $\eta_{\rm BR} := |\zeta/\Omega_\chi| \gg 1$, where only specific three-qubit transitions are allowed under resonant global driving.

To realize universal quantum computation, single- and two-qubit gates must be implemented. Single-qubit rotations are achieved by aligning the ICC with a crossed $B$- or $C$-type qubit and applying a sequence of global pulses tuned to generate effective local unitaries (see App.~\ref{app:ugc}).  Two-qubit entangling gates are performed when the ICC overlaps with a crossed $A$-type qubit bridging two rows, enabling a conditional phase gate between neighboring computational qubits (see App.~\ref{app:onetwogates} for the specific pulse sequences). The system's topology, together with the tunable pulse sequences and crossed elements, ensures that both gate types can be implemented efficiently within the globally controlled architecture (see App.~\ref{app:model} for gate definitions).

\section{Effect of disorder in superconducting architectures}
\label{sec:disordermain}

We analyze the impact of static inhomogeneities in the ladder quantum architecture introduced in Sec.~\ref{sec:intro-global-main}, focusing on how imperfections in superconducting qubit parameters degrade the fidelity of quantum information flow and gate operations. Specifically, we consider deviations in the local qubit frequencies $\omega_i$ and the nearest-neighbor longitudinal couplings $\zeta_{i,j}$, modeled as Gaussian-distributed random variables centered around nominal values $\bar\omega$ and $\bar\zeta$ with relative disorder percentage $\epsilon$. The system we studied has $N=2$ computational qubits, organized in a ladder formed by 15 \textit{physical} qubits. In simulations we used the Rabi frequencies $\Omega_{\chi} = 10$\,MHz, coupling strength $\bar\zeta = 2\pi \times 200$\,MHz, and nominal qubit frequencies $\bar\omega = 2\pi \times 7$\,GHz, with interaction strength-to-drive ratio $\eta_{\rm BR} = 20$. Here, we present a brief account of these numerical investigations, while an in-depth analysis is presented in App.~\ref{app:inhomogenjul}.

Disorder is introduced as follows: for each qubit $i$, the internal frequency and its nearest-neighboring coupling are sampled from the distributions 
\begin{align}
    \omega_i &\sim \mathcal{N}(\bar{\omega}, \epsilon\bar{\omega}), \\
    \zeta_{i,j} &\sim \mathcal{N}(\bar{\zeta}, \epsilon\bar{\zeta}),
\end{align}
such that $99.7\%$ of disorder realizations lie within $\omega_i \in \bar{\omega}(1 \pm \epsilon)$. Above, $\mathcal{N}(\mu,\sigma)$ refers to the normal Gaussian distribution with mean $\mu$ and standard deviation $\sigma$. We evaluate the performance of the protocol by computing the state fidelity, which measures the overlap between the disorder-averaged final state and the ideal, disorder-free target state. High fidelities indicate that the control pulses effectively mitigate the detrimental effects of static disorder, preserving the intended quantum evolution across different realizations. In contrast, a strong decay in fidelity signals reduced coherence transfer and degraded control performance due to frequency and coupling inhomogeneities.
 Specifically, we quantify the effect of disorder using the expression for the state fidelity~\cite{Nielsen2010},
\begin{equation}
    \mathcal{F}(p, \phi) := \big|\langle \Psi_{\mathrm{target}}(p,\phi) | \Psi_{\rm final}(p,\phi) \rangle\big|,
\end{equation}
between the ideal and actual many-body final states after propagating an initial ICC state of the form 
\begin{equation}
\vert \psi_{\mathrm{initial}}(p,\phi) \rangle = (\sqrt{1-p} \ket{0} + \sqrt{p}e^{i\phi} \ket{1}) \otimes  \ket{0}.
\end{equation}
across the ladder using global pulses.  Further modeling details and numerical methods are discussed in App.~\ref{app:model}. Above, $0\leq p\leq 1$ parameterizes the coherence of one of the qubits, with $p=0$ corresponding to the initial $|0\rangle$ state while $p=1$ corresponds to $|1\rangle$. We explore via numerical simulations how disorder affects the ICC movement as a function of $p$.

Simulations are performed using tensor network methods based on the \texttt{ITensor} library~\cite{ITensor, ITensor2}, where the ladder was mapped to an effective 1D model with long-range couplings. Time evolution is performed using the time-dependent variational principle (TDVP) combined with Krylov subspace methods~\cite{schollwock2005, schollwock2011, haegeman2011, yang2020}. Disorder averages are computed over up to $N_{\rm samples} = 300$ randomly sampled realizations, ensuring convergence with respect to bond dimension and time-step. Simulations employ the rotating frame together with the rotating wave approximation (RF+RWA), justified in the strong-interaction regime $\eta_{\rm BR} \gg 1$.

\begin{figure}[t]
\centering
\includegraphics[width=1.0\columnwidth]{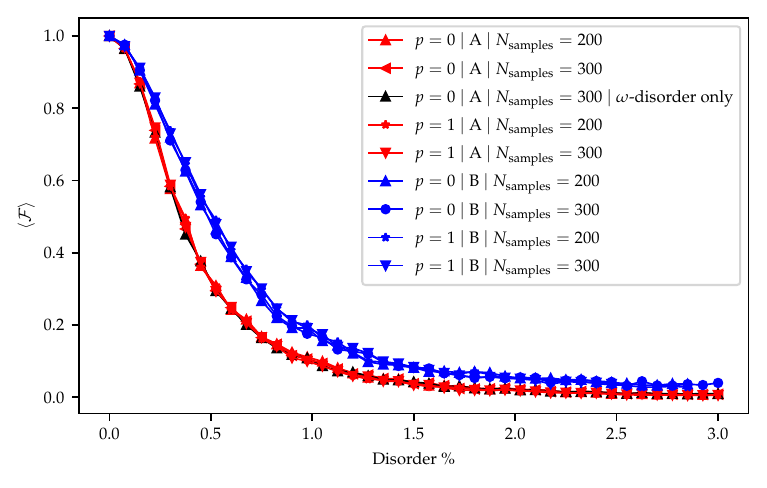}
\caption{
\label{fig:both}
\textbf{Fidelity w.r.t.~disorder percentage:}~$\mathcal{F}(p,0) = \vert \langle \Psi_{\mathrm{target}}(p,0) | \Psi_{\mathrm{final}}(p,0) \rangle \vert$. Fidelity $\mathcal{F}(p, 0)$ as a function of static disorder percentage, computed using the RF+RWA drive Hamiltonian for $N = 2$ qubits and $\eta_{\rm BR} = 20$.  Disorder is both in qubit frequencies ($\omega$) and couplings ($\zeta$), modeled as Gaussian noise with standard deviation $\epsilon \times \bar{\omega}$ and $\epsilon \times \bar{\zeta}$, respectively. The initial state is prepared on an $A$-type (solid blue) or $B$-type (solid green) ICC as $\vert \psi_{\mathrm{initial}} \rangle$. The solid pink line shows the fidelity under frequency disorder only. Results are averaged over $N_{\mathrm{samples}} \in \{200, 300\}$ disorder realizations. Due to the large energy scale separation, only the relative disorder percentage $\epsilon$ is plotted on the $x$-axis. ICC translation is performed using simultaneous $\pi$-pulses on $B$- and $C$-type qubits.
}
\end{figure}

Figure~\ref{fig:both} presents the fidelity $\mathcal{F}(p,0)$ as a function of the disorder percentage $\epsilon$ for inhomogeneities in $\omega$ and in $\zeta$. As we can see, the fidelity degrades significantly with increasing $\epsilon$, with notably lower performance for $A$-type ICC initialization due to the longer sequence required to complete a full shift operation. In this plot, the pink curve represents the fidelity when only disorder in the resonant frequency is applied. We find that this curve perfectly matches the one in blue, where also the disorder in $\zeta$ is present, meaning that the $\omega$-disorder is the dominant source of infidelity.
Indeed, Fig.~\ref{fig:zeta} shows that the disorder in $\zeta$ also reduces the fidelity, but to a lesser extent, consistent with the hierarchy $\omega \gg \zeta$ in energy scales. In Fig. \ref{fig:omega} in App.~\ref{app:inhomogenjul}, the fidelity against the presence of only $\omega$-disorder is shown.

Thus, our numerical results highlight that (i) disorder in the qubit frequency $\omega$ plays a more critical role than disorder in the coupling $\zeta$, and (ii) even modest inhomogeneities can lead to large reductions in fidelity if no compensation mechanisms are employed. This motivates the pulse optimization techniques discussed in Sec.~\ref{sec:pulse-optimization-main}, which are essential for restoring high-fidelity information flow in globally controlled superconducting systems.

\section{Overcoming disorder: pulse optimization}\label{sec:pulse-optimization-main}
A possible solution to mitigate the disorder problem is to develop improved pulse schemes for application to the QPU. To identify such sequences, one can employ quantum optimal control strategies. In particular, this work focuses on one of the most widely used methods, the GRadient Ascent Pulse Engineering (\texttt{GRAPE}) algorithm \cite{Khaneja2005_grape}. However, it is worth stressing that future work will focus on other schemes, as for instance stochastic gradient descent~\cite{ferrie,turinici}, the Krotov algorithm~\cite{krotov,goerz}, and various reinforcement learning approaches and their extensions~\cite{august,bukov}, as well as non-gradient-based methods such as the Chopped Random Basis (CRAB) algorithm~\cite{monta1,monta2} and the Nelder–Mead simplex method~\cite{kelly}.
\subsection{The \texttt{GRAPE} algorithm}\label{sec:grape}
\paragraph{Introduction to \texttt{GRAPE}.}
In this section, we provide a brief overview of the \texttt{GRAPE} algorithm. We begin with a quantum system described by a time-dependent Hamiltonian of the following form:
\begin{equation}
\hat H(t) = \hat H_0 + \sum_{j=1}^N u_j(t)\hat H_j\,.
    \label{grape Ham}
\end{equation}

\begin{figure}[h!]
\centering
\includegraphics[width=1.0\columnwidth]{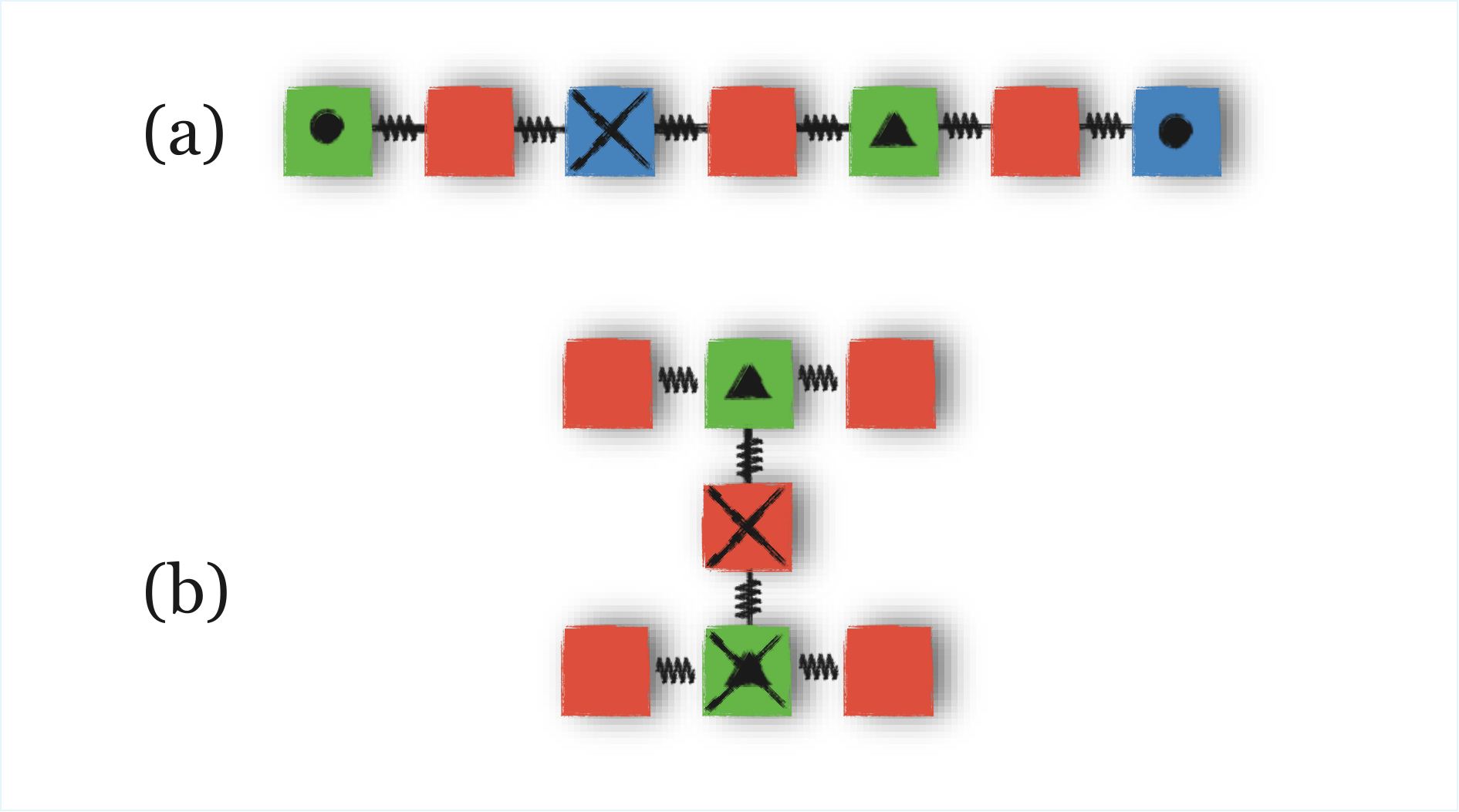}
\caption{
\label{fig:grape architecture}
\textbf{Portions of the QPU simulated in the \texttt{GRAPE} setting.}
This figure illustrates the two architectures used for our \texttt{GRAPE} simulations. In the upper part (a), a single row of 7 qubits is shown, representing an $N=1$ QPU capable of executing both Information Flow protocols and single-qubit gates. In the lower part (b), the architecture for implementing a two-qubit gate is depicted. The two data qubits are shown in green, while the red crossed intrarow qubit is used to perform a Controlled-Z gate between them, following our protocol. Note that this portion of the QPU is quite similar to a 7 qubits IBM architecture called \textit{ibm\_nairobi}~\cite{ibm_nairobi}.}
\end{figure}

Here, $\hat H_0$ is a static Hamiltonian representing the time-independent part of the system. We then consider $N$ different controls, each associated with a time-independent Hamiltonian $\hat H_j$, while their time dependence is captured by the $N$ functions $u_j(t)$, referred to as control amplitudes. The objective of the \texttt{GRAPE} algorithm is to determine the optimal $u_j(t)$ such that the system's dynamics under $\hat H(t)$ yields a desired target state. In general, obtaining analytical solutions for time-dependent Hamiltonians is not feasible; therefore, a piecewise constant approximation of the control amplitudes is employed. The total evolution time $T$ is divided into $M$ time slots, which may have different durations, within which each control amplitude is assumed to remain constant. Under this approximation, the Hamiltonian \eqref{grape Ham} can be expressed as
\begin{equation}
\hat H(t_k) = \hat H_0 + \sum_{j=1}^N u_{jk}\hat H_j\,,
    \label{grape Ham discrete}
\end{equation}
where $t_k$ is the evolution time at the beginning of the $k$-th time slot, and the control amplitudes are represented by an $N \times M$ matrix with components $u_{jk}$. In this framework, the time-evolution propagator for the $k$-th time slot is given by $\hat X_k = \exp[-i\hat H(t_k)\Delta t_k]$, where $\Delta t_k$ ($\hbar$ fixed to unity in what follows) is the duration of the time slot. Consequently, the evolution up to any time slot $k$ is obtained using the propagator
\begin{equation}
\hat X(t_k) = \hat X_k \hat X_{k-1}\cdots \hat X_1 \hat X_0\,.
    \label{grape propagator}
\end{equation}
To complete the setup, we must define the initial and final points of the dynamics, namely an initial state $X_0=\ket{\psi_0}$ or unitary operator $\hat X_0=\hat U_0$, and a target state $X_{\rm target}=\ket{\psi_1}$ or target unitary $\hat X_{\rm target} = \hat U_{\rm target}$. The objective of \texttt{GRAPE} is to determine the optimal $N \times M$ parameters that minimize a \textit{figure of merit} or \textit{cost function} $\mathcal{E}$ quantifying how closely the evolution under the Hamiltonian \eqref{grape Ham discrete} matches the target. This minimization is achieved by following the gradient with respect to the control amplitudes $u_{jk}$.
\paragraph{Mapping our setting to a \texttt{GRAPE} algorithm.}
To apply the \texttt{GRAPE} algorithm to our problem, we must map our system onto the framework required by \texttt{GRAPE}. The Hamiltonian terms can be mapped straightforwardly. Considering the Hamiltonian of our system~\eqref{eq:RWARFhammain}, the second term corresponds to $\hat H_0$ in Eq.~\eqref{grape Ham discrete}. The specific form of the first term in Eq.~\eqref{eq:RWARFhammain} arises from our modeling of the drive Hamiltonian as in Eq.~\eqref{eq:HdriveMain}. However, it can be shown that, in practical experimental settings, independent channels for the $\hat{\sigma}^{(x)}$ and $\hat{\sigma}^{(y)}$ operators are available (see, for example, Eq.~(2.8) in Ref.~\cite{Gambetta2011_drive}). Therefore, for each control line, we can consider two separate drive operators. Since, in our global control setting, the number of drive lines equals the number of species $\mathcal{S} =\{A,B, C\}$, the total number of control amplitudes is $N=6$, and the control term in Eq.~\eqref{grape Ham discrete} can be expressed as
\begin{equation}
\sum_j u_{jk}\hat H_j \equiv \sum_{\chi \in \mathcal{S}}\left(\Omega_{k,x}^{\chi}\hat{\sigma}^{(x)}_{\chi} + \Omega_{k,y}^{\chi}\hat{\sigma}^{(y)}_{\chi}\right)\,,
    \label{our control grape}
\end{equation}
where $\hat{\sigma}^{(\alpha)}_{\chi}$ indicates that the operator $\hat\sigma^{(\alpha)}$ is applied to all qubits belonging to species $\chi$. The number of time slots $M$ is not fixed \textit{a priori} and can be treated as a hyperparameter. However, since we aim to consider an experimentally feasible setting, we impose that the rate of change of the control amplitudes must be bounded by experimental constraints. In particular, state-of-the-art qubit controllers allow the amplitude to be updated once every $0.5$ ns \cite{zurich_instr}. Therefore, in our numerical simulations, we choose the value of $M$ to satisfy this constraint.
\paragraph{Defining the cost function: The operator trace-distance.}
To complete the \texttt{GRAPE} setup, we define the cost function and the target. Since we require the pulse sequence to be independent of the specific state of the QPU or the position of the ICC, we select as the target the unitary operator that implements a given operation, as defined in Eq.~\eqref{operation unitary}. We know how to combine such unitaries $\hat U^{(\ell)}_\chi$ to realize all the necessary logical operations (single- and two-qubit gates) as well as the operation that moves the interface across the QPU, as described in Eq.~\eqref{operation move}. Therefore, the initial unitary for the \texttt{GRAPE} algorithm is the identity, $\hat X_0 = \openone$, while the target unitary is the desired operation to be achieved through the optimal pulse sequence, $\hat X_{\rm target} = \hat U$. As the cost function $\mathcal E$, we use the trace distance between the target unitary and the result of the pulse sequence.
\begin{equation}
\mathcal E \equiv 1 - \frac{1}{d}\left|\mathrm{Tr}\big(\hat X(t_k)\hat X_{\rm target}\big)\right|\,,
    \label{cost function}
\end{equation}
with $d$ the Hilbert dimension of the system.
\paragraph{Numerical parameters.}
Finally, we present the numerical parameters used for all \texttt{GRAPE} runs. All numerical simulations are performed using the libraries provided in \texttt{QuTiP}~\cite{lambert2024qutip5quantumtoolbox}. Because pulse optimization is computationally demanding, we restrict the optimization to a small portion of the previously described QPU, reducing the analysis to 7 qubits. This number of qubits is the minimum required to perform all the relevant operations necessary for the QPU to function as a universal quantum computer. By arranging 7 qubits in a row, we can simulate both the moving interface protocol and single-qubit gates, as shown in Fig.~\ref{fig:grape architecture} (a). In particular, we focus on the Hadamard gate. Then, by arranging the qubits in a “reversed H” configuration (see Fig.~\ref{fig:grape architecture} (b)), we can implement a two-qubit gate, specifically a Controlled-Z (CZ) gate. From the disorder analysis, we observed that the dominant contribution to fidelity loss arises from inhomogeneities in the resonant frequency; therefore, we focus solely on this effect. In particular, for state-of-the-art transmon qubits, the spread in resonant frequency is approximately $2\%$ of the nominal value~\cite{Hertzberg2021_gauss_spread}. We thus use this value as the variance of the Gaussian distribution describing the static disorder introduced in the QPU. Note that this level of disorder corresponds to regimes where fidelities are close to zero (see Fig.~\ref{fig:omega}). For simplicity, we assume that all qubits have the same resonant frequency of $7$~GHz. For the ZZ coupling ($\zeta$) and the Rabi frequency $\Omega_\chi$, we choose $200$~MHz and $10$~MHz, respectively, yielding a favorable value for the blockade regime parameter, $\eta_{\rm BR} = 20$. In subsequent analyses, we increase the Rabi frequency to explore faster pulse sequences, thereby reducing $\eta_{\rm BR}$ to $5$.
Regarding disorder, we perform \texttt{GRAPE} iterations for a specific sample of resonant-frequency spread, reflecting the realistic possibility of experimentally calibrating and determining all physical parameters of the QPU. In other words, we do not aim to find a pulse sequence that works for all possible disorder configurations but rather for a single specific realization of disorder. We then conduct further analysis to assess the robustness of the optimized pulse sequence against variations and fluctuations in this particular disorder realization.

\subsection{Information flow}\label{sec:PO info flow}
In this section, we present our results. As described in Eq.~\eqref{cost function}, the fidelity during the \texttt{GRAPE} algorithm is evaluated using the trace distance between the desired operator and the one obtained by applying a given pulse sequence. To verify the validity of the pulse sequence produced by the \texttt{GRAPE} routine, we compute the state fidelity, defined as the overlap between the desired final state of the QPU and the state obtained from the time evolution of the initial state, $\mathcal{F}(t) = |\langle\Psi_{\rm target}|\Psi(t)\rangle|$. To obtain a global estimate of the pulse quality, we average this fidelity over a set of initial states of the computational qubit, denoted as $\ket{\psi_{\mathrm{initial}}(p,\phi)}$. The resulting averaged fidelity is indicated as $\overline{\mathcal{F}}$.

For the Information Flow (I.F.) protocol, the results are summarized in Table~\ref{IF results}. We conducted the analysis in both the strong blockade regime $\eta_{\rm BR} = 20$ and the weaker regime $\eta_{\rm BR} = 5$. The latter corresponds to higher Rabi frequencies for the drive, enabling faster pulse sequences. We examined protocols that move the ICC starting from both $A$-type and $B$-type qubits and obtained comparable results in both cases. Furthermore, we verified that the same pulse sequence performs effectively when moving the ICC from right to left or left to right. 

An example of optimal pulse sequence is depicted in Fig.~\ref{fig:pulse example} in App.~\ref{app:pulse example}.
\begin{table}[h!]
\begin{tabular}{|c|c|c|c|c|}
\hline
$\eta_{\rm BR}$ & Initial & Time ($ns$) & $\overline{\mathcal{F}}$ without \texttt{GRAPE} & $\overline{\mathcal{F}}$ with \texttt{GRAPE} \\ \hline
20     & $B$               & 2800        & $0.112\pm0.008$                          & $0.984\pm 0.007$                      \\ \hline
20     & $A$               & 2800        & $0.045\pm0.005$                          & $0.974\pm 0.008$                      \\ \hline
5      & $B$               & 700         & $0.028\pm0.007$                          & $0.989\pm 0.001$                      \\ \hline
5      & $A$               & 700         & $0.019\pm0.006$                          & $0.9945\pm 0.0004$                    \\ \hline
\end{tabular}
\caption{\textbf{\texttt{GRAPE} results for Information Flow.} We investigate two blockade regimes, $\eta_{\rm BR} = 20$ and $\eta_{\rm BR} = 5$, as well as protocols that move the ICC starting from either an $A$-type qubit or a $B$-type qubit. The table also reports the time required to perform each operation to the first significant digit, based on the numerical parameters presented in the text. It is evident that the \texttt{GRAPE} algorithm significantly outperforms the naive protocol in the presence of disorder.}
\label{IF results}
\end{table}

\begin{figure}[t]
\centering
\includegraphics[width=1.0\columnwidth]{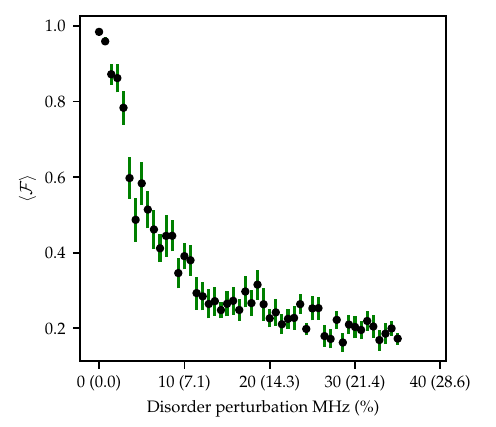}
\caption{
\label{fig:resilience}
\textbf{Resilience w.r.t.~disorder perturbation.} The $x$-axis represents the magnitude of the additional disorder applied to each resonant frequency (percentage of the perturbation w.r.t the static disorder $\delta\omega$ is in parentheses). For each point, we sample multiple disorder realizations and average the resulting fidelity. It is evident that even a small perturbation, on the order of $1$~MHz, leads to a dramatic drop in fidelity. Therefore, achieving high-resolution measurements of the static disorder within the chip is essential to obtain good fidelities.}
\end{figure}

\paragraph{Resilience to disorder perturbations.}
It is interesting to analyze the resilience of the \texttt{GRAPE} protocol to disorder perturbations. To this end, we repeat the previous analysis by adding a random contribution to the resonant frequency of each qubit. In this setting, each qubit has a frequency $\omega = \bar{\omega}+\delta\omega+\delta\omega_1$, where $\bar{\omega}$ is the nominal resonant frequency, $\delta\omega$ is the static disorder already included in the \texttt{GRAPE} setup, and $\delta\omega_1$ is an additional disorder term sampled from a Gaussian distribution centered at 0, with a standard deviation proportional to that of $\delta\omega$. For each value of the spread of $\delta\omega_1$, we generate multiple disorder samples and average the resulting fidelity. The results for a representative pulse sequence are shown in Fig.~\ref{fig:resilience}. It is evident that \texttt{GRAPE} performance degrades significantly under small perturbations, suggesting that high-precision measurements of the static disorder are necessary to achieve good fidelities.

\subsection{Single-qubit and two-qubit gates}\label{sec:PO 1 and 2 q gates}
\begin{figure}
\centering
\includegraphics[width=1.0\columnwidth]{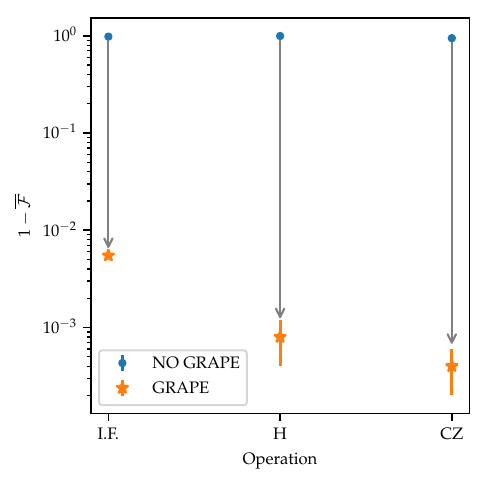}
\caption{
\label{fig:summary}
\textbf{Summary plot.} This plot shows the best result obtained for each operation. For improved resolution, we plot the infidelity $1-\overline{\mathcal{F}}$  on a logarithmic scale. The blue dots represent the fidelity error for the naive protocol in the presence of disorder, while the orange stars correspond to the error obtained using the pulse sequences optimized with the \texttt{GRAPE} algorithm. It is evident that \texttt{GRAPE} effectively mitigates the impact of static disorder, at least within our setting. Interestingly, the best fidelity is achieved for the two-qubit gate. This is because the protocol for this gate requires fewer pulses and only one species ($A$) is driven, making it less susceptible to static disorder. }
\end{figure}

In this section, we present the results obtained using \texttt{GRAPE} to optimize pulse sequences for single-qubit and two-qubit gates. Specifically, we focus on the Hadamard gate and the CZ gate. Once again, we achieve fidelities close to unity, as reported in Table \ref{gate results}, whereas the naive protocol performs poorly in the presence of static disorder. We summarize the best fidelities achieved with the \texttt{GRAPE} algorithm in Fig.~\ref{fig:summary}, using  QuTip \cite{lambert2024qutip5quantumtoolbox}.

Beyond the QuTiP-based GRAPE optimizations, we have also developed a custom MPS-based pulse optimization framework to extend scalability to larger quantum processors. While the previous simulations relied on exact state-vector methods limited to 7-qubit subsystems, the new approach leverages matrix product state (MPS) representations to efficiently capture the entanglement structure of the ladder architecture, enabling pulse optimization on significantly larger systems with high accuracy. This implementation integrates the GRAPE methodology with time-dependent variational principle (TDVP) propagation and Adam optimization, and performs \textit{state-based} rather than gate-based training by maximizing the average fidelity over a set of initial and target states. Control amplitudes are constrained to experimentally realistic nanosecond-scale resolutions. Using this framework, we successfully optimized a Hadamard gate on a 15-qubit ladder system with 2\% frequency disorder, demonstrating that high-fidelity pulse (on average 96\% for $p\in [0,1]$) design can be achieved not only with QuTiP but also with custom MPS-based numerical methods. The results are provided for completeness in App.~\ref{app:mps_grape}.

\begin{table}
\begin{tabular}{|c|c|c|c|c|}
\hline
Gate                     & $\eta_{\rm BR}$ & Time ($ns$) & $\overline{\mathcal{F}}$ without \texttt{GRAPE} & $\overline{\mathcal{F}}$ with \texttt{GRAPE} \\ \hline
H                        & 20     & 2500        & $0.075\pm0.005$                          & $0.985\pm 0.005$                      \\ \hline
H                        & 5      & 630         & $0.0052\pm0.0008$                        & $0.9992\pm 0.0004$                    \\ \hline
\multicolumn{1}{|c|}{CZ} & 20     & 940         & $0.055\pm0.004$                          & $0.9996 \pm 0.0002$                   \\ \hline
\multicolumn{1}{|c|}{CZ} & 5      & 235         & $0.062\pm0.008$                          & $0.987\pm0.003$                       \\ \hline
\end{tabular}
\caption{\textbf{GRAPE results for 1-qubit and 2-qubits gates.} Analogously to the I.F. protocol, we evaluate the performance of the \texttt{GRAPE} algorithm in two blockade regimes, $\eta_{\rm BR} = 20$ and $\eta_{\rm BR} = 5$, also reporting the corresponding operation times to the first significant digit. Once again, it is evident that the algorithm successfully identifies pulse sequences that achieve fidelities close to unity.}
\label{gate results}
\end{table}

\subsection{Reducing operation time}
In the previous sections, we have shown that the \texttt{GRAPE} algorithm can identify suitable pulse sequences to mitigate the effects of inhomogeneities in a globally controlled QPU. Furthermore,  as we show in this section, it can be used to find faster pulse sequences in global control settings as well. To this end, we attempt to reduce the gate time by half, as reported in Tables \ref{IF results} and \ref{gate results}. For the I.F. protocol and the Hadamard gate, we achieve promising fidelities with gate times close to $300$ ns. However, we were unable to shorten the duration of the CZ gate, likely because its execution time is already on the order of $200$ ns in the $\eta_{\rm BR} = 5$ case. We summarize these results in Table \ref{tab:reduced time results}. 

In addition to these results, we explored pulse compression using the MPS-based optimization framework introduced in the previous section. By exploiting the efficiency of the MPS representation and its ability to capture relevant entanglement features with reduced computational overhead, we were able to identify even shorter pulse sequences while maintaining high fidelities. In particular, for the Hadamard gate, we achieved gate durations below $250$~ns (160~ns, App.~\ref{app:mps_grape}) without significant degradation in performance, demonstrating that the MPS-based GRAPE approach can effectively balance speed and accuracy. These results highlight the potential of MPS-accelerated optimization to further reduce operation times beyond those obtained with standard QuTiP implementations, offering a promising pathway for scalable and time-efficient control in larger quantum processors.

\begin{table}[h!]
\begin{tabular}{|c|c|c|c|c|}
\hline
\multicolumn{1}{|l|}{Gate} & $\eta_{\rm BR}$ & Time (ns) & $\overline{\mathcal{F}}$ without \texttt{GRAPE} & $\overline{\mathcal{F}}$ with \texttt{GRAPE} \\ \hline
H                          & 20     & 1250        & $0.0035\pm0.0005$                        & $0.996\pm 0.001$                      \\ \hline
H                          & 5      & 315         & $0.055\pm0.006$                          & $0.982\pm 0.002$                      \\ \hline
I.F. ($A$)                   & 20     & 1415        & $0.195\pm0.008$                          & $0.994 \pm 0.002$                     \\ \hline
I.F. ($A$)                   & 5      & 350         & $0.112\pm0.005$                          & $0.9937\pm0.0005$                     \\ \hline
\end{tabular}
\caption{\textbf{GRAPE result for reduced operation times.} Here, we present the fidelities obtained for pulse sequences with operation times reduced by half compared to the naive sequences. In the best cases, we achieve very high fidelities with gate times close to $300$ ns. However, we were unable to shorten the duration of the two-qubit gate, as the naive protocol is already relatively fast.}
\label{tab:reduced time results}
\end{table}

\section{Conclusions}\label{sec:conclusion}

In this work, we have demonstrated that globally-driven superconducting quantum computing architectures, specifically the ladder-based design recently proposed in Ref.~\cite{menta2024globally}, can effectively overcome the detrimental effects of fabrication-induced static disorder through optimized control strategies. While in the case of global architectures based on Rydberg atoms \cite{cesa2023universal} such disorder is not present, solid state architectures typically suffer of it. As expected, our numerical simulations show that, in the absence of mitigation strategies at the level of pulses, even modest levels of disorder, particularly in the qubit resonance frequencies (on the order of $2\%$), can reduce quantum operation fidelities to nearly zero. This is especially true for information transport sequences involving $A$-type qubits, which require longer pulse trains. However, such disorder levels are fully compatible with typical fabrication variability in state-of-the-art transmon-based superconducting qubits.

Specifically, we have shown that by employing quantum optimal control techniques, particularly the \texttt{GRAPE} algorithm~\cite{Khaneja2005_grape}, it is possible to identify robust pulse sequences tailored to specific disorder realizations. A summary of these results is provided in Tab.~\ref{tab:reduced time results} and Fig.~\ref{fig:summary}. As discussed, these sequences not only restore high-fidelity performance, with fidelities exceeding $0.99$ for both single- and two-qubit gates, but also enable the reliable movement of logical qubits across the ladder via global operations.

Remarkably, this holds even when operation times are reduced by more than a factor of two, indicating that the \texttt{GRAPE} framework can also be leveraged to achieve faster gate implementations while remaining within practical experimental constraints (e.g., sub-nanosecond update rates on control electronics). It is important however to stress that such pulses can excite higher order states in the bosonic ladder of superconducting transmon qubits. 

Our findings highlight the dual utility of \texttt{GRAPE} optimization in globally controlled architectures. Firstly, it allows high-fidelity computation despite realistic levels of disorder in $\omega$ and $\zeta$, with a maximum limit value of $2\%$, see Figs.~\ref{fig:both} and \ref{fig:summary}. Secondly, it accelerates gate sequences without sacrificing performance. Furthermore, our resilience analysis shows that while optimized sequences are sensitive to additional disorder not accounted for in the optimization, high-precision calibration of qubit parameters remains a viable and effective strategy to ensure robustness.

These results support the viability of globally controlled superconducting quantum architectures as a scalable alternative to fully local control, particularly given their reduced wiring complexity and inherent hardware efficiency. While our simulations were limited to a minimal $N=2$ ladder instance, the approach is scalable in principle, and future studies may leverage tensor network methods to explore larger systems. Moreover, the extension of the current control framework to more realistic three-level models of transmon qubits (e.g., including leakage to the $|f\rangle$ level) represents a promising direction for future work, and to other pulse optimization schemes ~\cite{ferrie,turinici,krotov,goerz,august,bukov,monta1,monta2,kelly, hu2025universal}.

\acknowledgments
M.P. and V.G. are co-founders and shareholders of Planckian.  At the time of their contributions, authors affiliated with Planckian are either employees of Planckian or PhD students collaborating with Planckian.

The data that support the findings of this article are openly
available \cite{data}

\clearpage
\appendix
\onecolumngrid
\begin{center}
\Large \textbf{APPENDICES}
\end{center}

\section{The ladder architecture}\label{sec:app-global}

In this section, we briefly summarize the working principle of a globally-driven superconducting quantum computing architecture, referred to as the {\it ladder} architecture, recently proposed in Ref.~\cite{menta2024globally}. 

The architecture consists of a ladder-like arrangement of nearest-neighbor coupled superconducting qubits, categorized into three types: $A$ [red], $B$ [blue], and $C$ [green]. If the ladder geometry is made by $N$ rows then $2N+3$ is the number of columns. Considering the additional mediating qubit connecting two nearest-neighboring rows for a total of $N-1$ qubits, the final amount of physical qubits is $N_{\rm tot}=2N^2+4N-1$ such that $\mathcal{O}(N_{\rm tot}) = \mathcal{O}(N^2)$.

Each qubit type $\chi \in \mathcal{S} := \{A,B,C\}$ is characterized by a specific internal frequency $\omega_{\chi}$ and is driven by a corresponding control signal $V_{\chi}(t)$. The qubits are arranged in the following periodic pattern
$CABACABA\ldots$ and the system is controlled globally via three independent drive lines, which operate either simultaneously or in separate time windows. Each qubit, regardless of its type, is coupled to its intra-row nearest neighbors via a ZZ interaction with coupling strength $\zeta$. Additionally, some $C$- or $B$-type qubits are also ZZ-coupled to certain $A$-type qubits positioned between two horizontally neighboring qubit rows.  

On Fig.~\ref{fig:ladder}, the (minimal) $N=2$ ladder-lattice architecture is illustrated~\cite{menta2024globally}, which will be the focus of this study due to its reasonable number of physical qubits, $N_{\rm tot}=15$, making it suitable for numerical simulations. In the figure, qubits are represented as colored squares, while black springs and continuous colored lines indicate the ZZ interactions and drive lines, respectively. Black circles on the qubits in the first and last columns denote an internal frequency of $\omega_\chi - \zeta$, whereas qubits marked with black triangles have an internal frequency of $\omega_\chi + \zeta$. These frequency inhomogeneities arise from the number of connections a qubit has: one for qubits with circles, two for empty qubits, and three for qubits with triangles.

Another key feature of the architecture is the presence of ``crossed'' qubits, represented as crossed squares in Fig.~\ref{fig:ladder}. These qubits have a coupling strength with their control source, i.e.~the Rabi frequency, which is twice that of regular (non-crossed) qubits. Thus, qubits are categorized not only by type but also by whether they are regular or crossed:  $\chi = \{\chi^{\rm r}, \chi^{\times}\}$, such that $\Omega_{\chi^\times} = 2\Omega_{\chi^{\rm r}}$ for all $\chi \in \mathcal{S}$.
Here, $\Omega_{\chi}$ denotes the Rabi frequency, which will be introduced as a time-dependent control parameter associated with the driving signal $V_{\chi}(t)$ in the following.

We now describe how the system in Fig.~\ref{fig:ladder} functions as a quantum processing unit, following a constructive approach. To do so, we first define how computational qubits are encoded. Consider a single row of the 2D ladder: it encodes one computational qubit, which is localized on one of the $2N-1$ physical qubits forming the processing unit area of that row. As demonstrated in Refs.~\cite{cesa2023universal, menta2024globally}, the computational qubit ($|\psi\rangle = \alpha|g\rangle + \beta|e\rangle$) is encoded at the interface between two distinct phases of physical qubits: a left (right)-hand N\'eel phase ($|geg\rangle$) and a left (right)-hand ferromagnetic phase ($|ggg\rangle$). By applying global operations alternately to the three qubit types, the computational qubit can be coherently shifted in either direction along the row, see Fig.~\ref{fig:spectrum-flow}~(b). In this sense, the physical qubits serve as a dynamic platform that hosts and transports quantum information.  

Now, let us generalize this concept to a 2D ladder with $N$ rows, each corresponding to a computational qubit. Since the rows are largely decoupled, except at specific points, they can be controlled in parallel, with quantum information being vertically encoded along a column known as the “information-carrying column” (ICC) which remains positioned at the boundary between the N\'eel phase ($|geg\rangle^{\otimes N}$) and the ferromagnetic phase ($|ggg\rangle^{\otimes N}$). By designing an appropriate sequence of global operations, the ICC can be coherently moved to any position in the ladder, while maintaining phase coherence.  

To achieve universal quantum computation, it is necessary to implement single-qubit gates on individual computational qubits and two-qubit gates between adjacent computational qubits within the ICC. This is enabled by the introduction of crossed qubits, which serve as localized control elements despite the system being globally driven. Specifically, $N$ crossed qubits of type $B$ or $C$ allow for single-qubit operations, while $N-1$ crossed qubits of type $A$ mediate two-qubit interactions between neighboring computational qubits. As discussed in Refs.~\cite{menta2025building, menta2024globally, cioni2024conveyorbelt, cesa2023universal}, qubits with different Rabi frequencies are essential for enabling local control within a globally driven architecture. The configuration in Fig.~\ref{fig:ladder} for $N=2$ represents a minimal yet optimized setting that reduces the total number of physical qubits, $N_{\rm tot}$.  

In conclusion, this system constitutes a fully functional quantum processing unit.  

\subsection{Hamiltonian model}\label{app:model}
The Hamiltonian describing the entire system is given by  $\hat{H} := \hat{H}_0 + \hat{H}_{\rm drive}(t)$, where: 
\begin{equation}\label{eq:H0}
    \hat{H}_0 := \sum_{\chi \in \mathcal{S}} \sum_{i \in \chi} \frac{\hbar \omega_i}{2} \hat{\sigma}^{(z)}_{i}
    +  \sum_{\langle i,j \rangle} \frac{\hbar \zeta}{2} \hat{\sigma}^{(z)}_{i} \hat{\sigma}^{(z)}_{j},
\end{equation}  
describes the local energy contributions $\hbar \omega_i$ of superconducting qubits in the ladder and their always-on ZZ interactions between nearest-neighbor qubits $\langle i,j\rangle$. The time-dependent driving term is given by:
\begin{equation}
\label{eq:Hdrive} 
    \hat{H}_{\rm drive}(t) := \sum_{\chi \in \mathcal{S}} \sum_{i \in \chi} \hbar \Omega_\chi(t)\sin(\omega_{\mathrm{d},\chi}t + \phi_{\chi}(t)) \hat{\sigma}^{(y)}_i,
\end{equation}  
which represents the interaction induced by classical control fields. Here, $\omega_{\mathrm{d},\chi}$ is the oscillation frequency of the driving pulse $V_\chi(t)$, while $\Omega_\chi(t)$ and $\phi_\chi(t)$ denote the time-dependent Rabi frequency and phase of the control, respectively. Importantly, these control parameters are independent of the qubit index $i$, indicating that each control acts {\it globally} on all qubits of type $\chi$. Note that in the time intervals where the control $V_{\chi}(t)$ is active, we shall assume $\Omega_\chi(t)$ and $\phi_\chi(t)$ to be constant.
In Eqs.~\eqref{eq:H0} and \eqref{eq:Hdrive}, the operators $\hat{\sigma}_i^{(x,y,z)}$ denote the Pauli matrices acting on the Hilbert space of the $i$-th qubit, expressed in the local energy basis with the convention $\ket{g_i} := (0,1)^{\rm T}$, 
$\ket{e_i} := (1,0)^{\rm T}$.

For simplicity, Eq.~(\ref{eq:H0}) does not explicitly account for frequency shifts in certain qubits. Specifically, if $i$ corresponds to a $\chi=B,C$-type qubit marked with a black circle or triangle, its level spacing is modified as $\omega_i = \omega_{\chi}^{\bullet} := \omega_{\chi} - \zeta$ or $\omega_i = \omega_{\chi}^{\blacktriangle} := \omega_{\chi} + \zeta$, respectively, instead of $\omega_\chi$. Similarly, since each qubit type can be either regular or crossed, i.e., $\chi = \{\chi^{\rm r}, \chi^\times\}$ for all $\chi \in \mathcal{S}$, when $i$ corresponds to a crossed qubit, the Rabi frequency doubles: $\Omega_{\chi}(t) \to 2\Omega_\chi(t)$.

Taking these modifications into account, Eq.~\eqref{eq:H0} can be rewritten as $\hat{H}_0 := \hat{H}^{\rm bare}_0 + \hat{H}^{\bullet}_0 + \hat{H}^\blacktriangle_0$, where $\hat{H}^{\rm bare}_0$ represents the unmodified free Hamiltonian of Eq.~\eqref{eq:H0}, and:
\begin{equation}
    \hat{H}^{\bullet}_0 + \hat{H}^\blacktriangle_0 := \sum_{\chi \in \{B,C\}} \sum_{i \in \chi} \Bigg(\frac{\hbar \omega^\bullet_i}{2} \hat{\sigma}^{(z)}_{i} + \frac{\hbar \omega^{\blacktriangle}_i}{2} \hat{\sigma}^{(z)}_{i} \Bigg),
\end{equation}  
accounts for the local energy contributions of qubits with shifted frequencies.  
Similarly, the driving term can be rewritten as  $\hat{H}_{\rm drive}(t) := \hat{H}_{\rm drive}^{\rm r}(t) + \hat{H}^{\times}_{\rm drive}(t)$, where $\hat{H}^{\rm r}_{\rm drive}(t)$ corresponds to the driving term for regular qubits given by 
Eq.~\eqref{eq:Hdrive} with $\chi = \chi^{\rm r}$, while:  
\begin{equation}
    \hat{H}^{\times}_{\rm drive}(t) := \sum_{\chi \in \mathcal{S}} \sum_{i \in \chi^{\times}} \hbar 2\Omega_\chi(t)\sin(\omega_{\mathrm{d},\chi}t + \phi_{\chi}(t)) \hat{\sigma}^{(y)}_i,
\end{equation}  
describes the driving Hamiltonian for crossed qubits.  

Notice that the Rabi frequency can be expressed as $\Omega_{\chi}(t) = \mathcal{V} f_{\chi}(t)$, where $f_{\chi}(t)$ is an envelope function, and $\mathcal{V} $ represents the coupling strength between the qubits and the waveguide $V_{\chi}$. This coupling strength can be set during the fabrication process such that certain qubits, specifically the crossed ones, exhibit a doubled coupling strength, i.e. $2\mathcal{V}$, compared to regular qubits. Importantly, despite this modification, the control remains {\it global}~\cite{menta2025building}.  \\

\begin{figure*}[t!]
\centering
\includegraphics[width=0.85\textwidth]{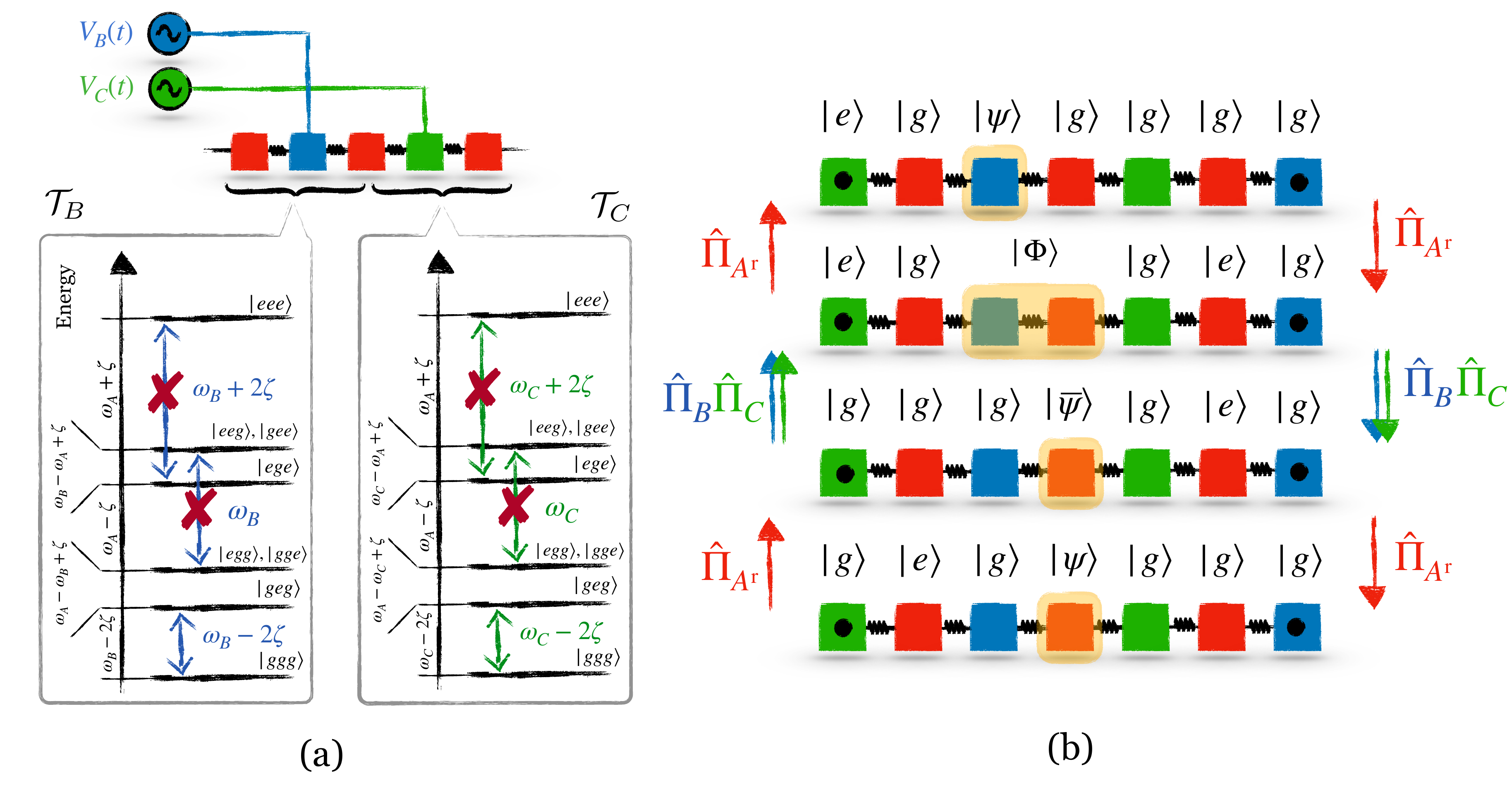}
\caption{
\label{fig:spectrum-flow}
\textbf{Illustration of quantum information dynamics.} (a) Energy spectrum of five qubits, arranged as $ABACA$ and coupled via nearest-neighbor ZZ interactions, in the time window $\mathcal{T}_{B,C} = \mathcal{T}_B \cap \mathcal{T}_C$. A driving pulse with frequency $\omega_{\mathrm{d},\chi} = \omega_{\chi} - 2\zeta$, applied to the $B$- and $C$-type elements, can only activate transitions between the two lowest energy levels: $|ggggg\rangle$ and $|gegeg\rangle$. For simplicity, the spectrum in the figure is divided into two adjacent triplets that share a central $A$-type qubit.  (b) A single-step information flow of a computational qubit $|\psi\rangle$ in the absence of disorder along a single row ($N=1$) of the ladder. For simplicity, no crossed elements are shown. The unitary transformation inducing this collective motion is $\hat{U}_{\rm shift}= \hat{\Pi}_{A^{\rm r}}\hat{\Pi}_{B}\hat{\Pi}_{C}\hat{\Pi}_{A^{\rm r}}$.  
}
\end{figure*}

Let us now consider the total Hamiltonian $\hat{H}(t)$ transformed into a rotating reference frame at the drive frequency $\omega_{\rm d,\chi}$. This transformation is formally achieved by applying the unitary operator $\hat{U}_{\rm rf}(t) := \bigotimes_i e^{i\hat{\sigma}_i^{(z)} \omega_{\mathrm{d,}i}t/2}$.
Within the rotating wave approximation (RWA) and with the specific choice for the drive frequency, i.e. $\omega_{\mathrm{d,}i}=\omega_i - 2\zeta$, the transformed Hamiltonian simplifies to:
\begin{equation}
\hat{H}_{\text{rf}}(t) \simeq \sum_{\chi \in \mathcal{S}} \sum_{i \in \chi} \frac{\hbar \Omega_\chi(t)}{2} \Big[e^{i\phi_{\chi}(t)}\vert g_i \rangle \langle e_i \vert + {\rm h.c.} \Big] + \sum_{\langle i,j \rangle} 2 \hbar \zeta \vert e_i e_{j} \rangle \langle e_i e_{j} \vert.
\label{app:eqhamrf}
\end{equation}  
If the dimensionless parameter $\eta_{\rm BR} := \left| \zeta/\Omega_{\chi} \right| \gg 1$, i.e. if the ZZ coupling is the dominant energy scale, then $\hat{H}_{\text{rf}}(t)$ precisely emulates the effective Rydberg-blockade Hamiltonian utilized in Ref.~\cite{cesa2023universal}. It is important to highlight that this effective interaction is dynamically induced by the driving field. Specifically, the system must be driven at $\omega_{\mathrm{d,}i}=\omega_i - 2\zeta$ to selectively allow transitions between the triplet states $ |ggg\rangle \leftrightarrow |geg\rangle $.  
This mechanism can be intuitively visualized in the simplified case of a single-row dynamics involving a triplet of qubits of type $ABA$ (or equivalently $ACA$), as depicted in Fig.~\ref{fig:spectrum-flow}~(a). Since the global control fields $V_\chi(t)$ can be activated within distinct time windows, this argument extends to the entire row and, consequently, to the entire ladder structure.

\subsection{Universal global control}\label{app:ugc}
Moreover, during each time window $\mathcal{T}_{\chi}$, the control $V_{\chi}(t)$ is active and constant, while all other controls are set to zero. More precisely, in our analysis, we impose the following disjoint activation conditions: if $V_A(t)$ is active (i.e., nonzero), then both $V_B(t)$ and $V_C(t)$ must be null. Conversely, if either $V_B(t)$ or $V_C(t)$ (or both) are active, then $V_A(t)$ must be null. This can be formally expressed as:
\begin{equation} 
\begin{cases}
    V_A(t) \neq 0 \Longrightarrow V_B(t)=V_C(t)=0,
\\
    V_B(t) \neq 0 \text{ or } V_C(t) \neq 0 \Longrightarrow V_A(t) = 0.
\end{cases}
\end{equation}  
Thus, the global control remains {\it disjoint}, ensuring selective activation of different qubit groups while preventing interference between them.
Formally, this constraint allows us to divide the temporal evolution into a collection of disjoint time windows, ${\cal T}_{\chi_1}, {\cal T}_{\chi_2}, \dots, {\cal T}_{\chi_\ell}$, where $\chi_{\ell} \in \mathcal{S}$. Within each window, the Hamiltonian $\hat{H}_{\text{rf}}(t)$ in Eq.~\eqref{app:eqhamrf} can be treated as time-independent, i.e. $\hat{H}_{\rm rf}(t) = \hat{H}^{(\ell)}_{\rm rf}$, $\forall t \in \mathcal{T}_{\chi_{\ell}}$. Consequently, the time-ordered unitary evolution over the total time window $\mathcal{T}_{\rm tot} = \bigcup_\ell \mathcal{T}_{\chi_\ell}$ simplifies to:
\begin{align} 
\label{stringSUP} 
& \hat{U}_{\rm tot} := \overleftarrow{\exp}\left[ -\frac{i}{\hbar} \int_{{\cal T}_{\rm tot}} dt' \hat{H}_{\text{rf}}(t')\right] = \hat{U}^{(\ell)}_{\chi}\cdots \hat{U}^{(2)}_{\chi}\hat{U}_{\chi}^{(1)},
\end{align}
where $\overleftarrow{\exp}[ ... ]$ denotes the time-ordered exponential and $\hat{U}^{(\ell)} := \exp \big(-i\hat{H}^{(\ell)}_{\text{rf}} \tau_\ell / \hbar \big)$ represents the unitary evolution associated with the $\ell$-th time interval (with $\tau_\ell$ being its duration).  
In the strong interaction regime, where $\eta_{\rm BR} := \vert \zeta/\Omega^{(\ell)}_{\chi} \vert \gg 1$, transitions induced by the driving term in $\hat{H}^{(\ell)}_{\text{rf}}$ are suppressed whenever they involve initial or final states in which at least one of the interacting neighbors of the $\chi$-type qubits affected by the control is in the excited state. More precisely, we will show that in the limit $\eta_{\rm BR} \gg 1$, each element $\hat{U}^{(\ell)}$ in the sequence~\eqref{stringSUP} can be expressed in terms of controlled unitary gates as: 
\begin{eqnarray} 
\label{eq:Uell}
\hat{U}^{(\ell)}_{\chi} :=  \prod_{i\in\chi_{\ell}} \left[  \hat{\openone}_i \hat{Q}_{\langle i \rangle}+\hat{\mathbb{U}}^{(\ell)}_i \hat{P}_{\langle i \rangle}  \right], 
\label{ccuSUP}
\end{eqnarray} 
up to a global unitary gate that can be postponed to the end of the process, as it plays no role in the dynamics. The blockade condition is enforced by the projectors $\hat{P}_{\langle i \rangle}$ and $\hat{Q}_{\langle i \rangle} = \hat{\openone}_{\langle i \rangle} - \hat{P}_{\langle i \rangle}$, where $\hat{P}_{\langle i \rangle}$ projects onto the subspace in which none of the nearest-neighboring qubits of the $i$th qubit are excited. The orthogonal complement $\hat{Q}_{\langle i \rangle}$ ensures that excitations in neighboring qubits prevent transitions, thus implementing the blockade mechanism.  
Finally, $\hat{\mathbb{U}}^{(\ell)}_i$ in Eq.~\eqref{eq:Uell} represents the single-qubit unitary evolution induced by the Rabi term of the Hamiltonian $\hat{H}^{(\ell)}_{\text{rf}}$ on the $\chi$-type qubits, given by:  
\begin{eqnarray} 
\hat{\mathbb{U}}^{(\ell)}_i &:=&\exp\Big[ -i \frac{\Omega^{(\ell)}_{\chi}\tau_\ell}{2} 
\big(e^{i\phi^{(\ell)}_{\chi}}\vert g_i \rangle \langle e_i \vert + {\rm H.c.} \big)\Big] \\ 
  &=& \nonumber
\exp\Big[ -i \frac{\Omega^{(\ell)}_{\chi}\tau_\ell}{2}  \left(  \cos(\phi^{(\ell)}_{\chi})~\hat{\sigma}^{(x)}_i + \sin (\phi^{(\ell)}_{\chi})~\hat{\sigma}^{(y)}_i\right)\Big],
\end{eqnarray} 
where $(\Omega_{\chi}^{(\ell)}, \phi_{\chi}^{(\ell)})$ are the constant control parameters within the time window $\mathcal{T}_{\chi_{\ell}}$.
A convenient way to rewrite the operators $\hat{U}^{(\ell)}_{\chi}$ defined in Eq.~(\ref{ccuSUP}) is to introduce the parameters:
\begin{equation}
\label{eq:theta-vector}
    \begin{cases}
        \theta^{(\ell)} := \Omega^{(\ell)}_{\chi}\tau_\ell, \\ 
        \vec{n}_\perp^{(\ell)} := \left(\cos\phi^{(\ell)}_{\chi}, \sin\phi^{(\ell)}_{\chi},0\right),
    \end{cases}
\end{equation} 
such that:
\begin{equation}
    \hat{\mathbb{U}}^{(\ell)}_i = \exp\left[-\frac{i\theta^{(\ell)}}{2} \vec{\sigma}_i \cdot \vec{n}_\perp^{(\ell)}\right],
\end{equation} 
where $\vec{\sigma}_i := \big(\hat{\sigma}^{(x)}_i,\hat{\sigma}^{(y)}_i,\hat{\sigma}^{(z)}_i\big)$. \\

Recalling that the crossed elements of the $\chi$-type qubits have twice the Rabi frequency of the regular ones, we can write:   
\begin{align} 
\hat{U}^{(\ell)}_{\chi} &=  
\prod_{i\in\chi_{\ell}^{\rm r}} \left[  \hat{\openone}_i \hat{Q}_{\langle i \rangle}+\hat{\mathbb{R}}_i(\theta^{(\ell)},\vec{n}_\perp^{(\ell)}) \hat{P}_{\langle i \rangle}  \right] \nonumber \times \prod_{i\in\chi_{\ell}^{\times}} \left[  \hat{\openone}_i \hat{Q}_{\langle i \rangle}+\hat{\mathbb{R}}_i(2\theta^{(\ell)},\vec{n}_\perp^{(\ell)}) \hat{P}_{\langle i \rangle}  \right];  \nonumber \\ 
\hat{U}^{(\ell)}_{\chi} &:=  \hat{W}_{\chi}(\theta^{(\ell)}, \vec{n}_\perp^{(\ell)}; 2\theta^{(\ell)}, \vec{n}_\perp^{(\ell)}),
\label{operation unitary}
\end{align} 
where, defining $\chi^{\rm r}$ and $\chi^{\times}$ as the regular (non-crossed) and crossed subsets of $\chi$-type qubits, we introduce:
\begin{align}
\label{WTRANSF}
& \hat{W}_{\chi}(\theta',\vec{n}'; \theta'',\vec{n}'') := \hat{W}_{\chi^{\rm r}}(\theta',\vec{n}')  \hat{W}_{\chi^{\times}}(\theta'',\vec{n}''); \\
& \hat{W}_{\xi}(\theta,\vec{n}):= \prod_{i\in \xi} \left[  \hat{\openone}_i \hat{Q}_{\langle i \rangle} + \hat{\mathbb{R}}_i(\theta,\vec{n}) \hat{P}_{\langle i \rangle}  \right],
\end{align} 
with $\xi\in \{ \chi^{\rm r},\chi^{\times}\}$ and:
\begin{align} 
& \hat{\mathbb{R}}_i(\theta,\vec{n}) = \exp\left[ -\frac{i\theta}{2} \vec{n} \cdot \vec{\sigma}_i \right] = \cos \left( \frac{\theta}{2} \right)\hat{\openone}_i - i\sin \left( \frac{\theta}{2}\right) \vec{n} \cdot \vec{\sigma}_i, 
\end{align} 
corresponding to the single-qubit unitary rotation associated with the unit vector $\vec{n}$ and the angle $\theta$. \\

By composing individual global pulsed unitaries $\hat{U}^{(\ell)}$, the total unitary $\hat{U}_{\rm tot}$ implements a specific type of controlled unitary gate in the form:
\begin{equation} 
\label{fdfsdf} 
\hat{W}_{\chi}(\theta, \vec{n}_\perp; 2\theta, \vec{n}_\perp) = \hat{W}_{\chi^{\rm r}}(\theta, \vec{n}_\perp)  \;  \hat{W}_{\chi^{\times}}(2\theta, \vec{n}_\perp),
\end{equation} 
which acts simultaneously on $\chi^{\rm r}$ and $\chi^{\times}$, inducing single-qubit rotations around a
generic unit vector $\vec{n}_\perp$ in the $xy$-plane (i.e., $\vec{z} \cdot \vec{n}_\perp = 0$) with correlated angles $\theta$ and $2\theta$. 
By combining sequences of these special pulse types~(\ref{stringSUP}) generated by the classical controls of the ladder, one can produce any transformation $\hat{W}_{\chi}(\theta',\vec{n}'; \theta'',\vec{n}'')$, where $\theta'$, $\theta''$, and $\vec{n}'$, $\vec{n}''$ are arbitrarily chosen. It is straightforward to show that, by separately applying Euler’s theorem to both $W_{\chi^{\rm r}}$ and $W_{\chi^{\times}}$, and then concatenating them, we can implement any arbitrary transformation of the form $\hat{W}_{\chi}(\theta',\vec{n}'; \theta'',\vec{n}'')$, see Ref.~\cite{menta2024globally} for further details.

\subsection{Information flow}\label{app:infoflow}
Under $e$-$e$ blockade condition, globally-controlled evolutions can be engineered; the latter enabling universal quantum computation. The essential features are: (i) the ability to move the ICC at any position on the ladder including the last column of the read-out area where quantum measurements can be performed at the end of the computation; (ii) the possibility to implement arbitrary single-qubit gates on the $B$- or $C$-type crossed element column where the ICC has to be previously located, while leaving the rest of the qubits invariant; (iii) the ability to activate a non-trivial two-qubit entangling gate, e.g. a CZ operation, between the $i$-th and $(i+1)$-th computational qubits when the ICC is located in the $B$- or $C$-type column containing the $A$-type crossed qubit; the latter acting as a coupler between the $i$-th and $(i+1)$-th rows of the ladder architecture. \\

We discuss here the task (i), namely the quantum information flow. Mathematically, this translates into: 
\begin{eqnarray}
\hat{U}^k_{\rm shift} \ket{\Psi; j} =  \ket{\Psi; j+k},
\label{operation move}
\end{eqnarray} 
up to an irrelevant global phase. $j$ and $k$ refer here to the initial and final position of the ICC whose associated quantum state is denoted by $\ket{\Psi}$. 
The real-time unitary operator $\hat{U}_{\rm shift}$ is given by: 
\begin{equation}
 \label{equshiftrev} 
   \hat{U}_{\rm shift} = \hat{\Pi}_{A^{\rm r}}\hat{\Pi}_{B}\hat{\Pi}_{C}\hat{\Pi}_{A^{\rm r}},
\end{equation}
where $\Pi_{\chi^\xi}$ denotes (global) pulses on all the $\chi^\xi$-type qubits with $\xi \in \{\rm r, \times\}$ and $\chi \in \mathcal{S}$. The explicit sequence of pulses is now presented. 
To engineer the global operation represented by the operator $\hat{\Pi}_{B}$ in Eq.~\eqref{equshiftrev}, the specific sequence is given by $\hat{\Pi}_{B} = \hat{U}^{(4)}_B\hat{U}^{(3)}_{B} \hat{U}^{(2)}_{B}\hat{U}^{(1)}_{B}$ where each global unitary is defined as (a) $\hat{U}^{(1)}_{B}$ is induced by fixing the global phase to $\phi_B^{(1)} = 0$ and letting the system evolve with a duration $\tau_1= \frac{3\pi}{4}\Omega_B^{-1}$; (b) 
$\hat{U}^{(2)}_{B}$ is characterized by the phase  $\phi^{(2)}_B = \pi/2$ and a duration $\tau_2= \pi\Omega_B^{-1}$; (c) $\hat{U}^{(3)}_{B}$ by $\phi^{(3)}_B = \pi$ and $\tau_3 = \frac{\pi}{4}\Omega_B^{-1}$, (d)
$\hat{U}^{(4)}_{B}$ by $\phi^{(4)}_B = -\pi/2$ and $\tau_4= \pi\Omega_B^{-1}$. We recall that, by convention, $\Omega_B$ denotes the Rabi pulsation associated to the regular $B$-type qubits. A similar sequence also applies to $\hat{\Pi}_{C}$. 
For $\hat{\Pi}_{A^{\rm r}}$, the $\pi$-pulses only act on the regular elements of the $A$-type qubits and are characterized by the following four-step sequence:
$\hat{\Pi}_{A^{\rm r}} = \hat{U}^{(4)}_{A}\hat{U}^{(3)}_{A} \hat{U}^{(2)}_{A}\hat{U}^{(1)}_{A}$, where (a) $\hat{U}^{(1)}_{A}$ is characterized by the phase $\phi_A^{(1)} = 0$ and the duration $\tau_1= \frac{\pi}{2} \Omega_A^{-1}$; 
(b) $\hat{U}^{(2)}_{A}$ by $\phi_A^{(2)} = \pi/2$ and $\tau_2 = \pi \Omega_A^{-1}$; (c) $\hat{U}^{(3)}_{A}$ by $\phi_A^{(3)} = \pi$ and $\tau_3 = \frac{\pi}{2} \Omega_A^{-1}$, (d) $\hat{U}^{(4)}_{A}$ by $\phi_A^{(4)} = -\pi/2$ and $\tau_4= \pi \Omega_A^{-1}$. $\Omega_A$ corresponds to the Rabi frequency of the regular $A$-type qubits.

\subsection{One- and two-qubit gates}\label{app:onetwogates}
We outline the implementation of one- and two-qubit gates.\\

For one-qubit gates on a $\chi$-type column ($\chi \in \{B,C\}$), we use the sequence:
\begin{eqnarray}
\hat{U}_{\chi,1} &=& \hat{W}_{A}(\pi, \vec{z}, 0, \vec{u}) \hat{W}_{\chi}(0, \vec{u}, \theta/2, -\vec{n}) \times \hat{W}_{A}(\pi, \vec{z}, 0, \vec{u}) \hat{W}_{\chi}(0, \vec{u}, \theta/2, \vec{n}),
\end{eqnarray}
where $\vec{n} \cdot \vec{z} = 0$, $\vec{u} \in \{\vec{x},\vec{y},\vec{z}\}$, and $\theta \in [0,2\pi[$. Using Eq.~\eqref{WTRANSF}, this simplifies to:
\begin{equation}
\hat{U}_{\chi,1} = \hat{Z}_{A^{\rm r}} \hat{W}_{\chi^{\times}}(\theta/2,-\vec{n}) \hat{Z}_{A^{\rm r}} \hat{W}_{\chi^{\times}}(\theta/2,\vec{n}),
\end{equation}
with $\hat{Z}_{A^{\rm r}} = \hat{W}_{A^{\rm r}}(\pi, \vec{z})$. Acting on $|\Psi; j\rangle$, this sequence implements the single-qubit rotation:
\begin{equation}
\hat{\mathbb{R}}(\theta,\vec{n}) = \hat{\sigma}^z \hat{\mathbb{R}}(\theta/2,-\vec{n}) \hat{\sigma}^z \hat{\mathbb{R}}(\theta/2,\vec{n}),
\end{equation}
which remains valid for generic input states $|\Psi; j\rangle$.\\

For two-qubit gates, we align the ICC with an $A^{\times}$-type coupler and apply:
\begin{equation}
\hat{U}_{A,2} = \hat{Z}_{A^{\times}} = \hat{W}_{A}(0, \vec{u}, \pi, \vec{z}) = \hat{W}_{A^{\times}}(\pi, \vec{z}),
\end{equation}
which acts as a conditional phase (CZ) gate between crossed qubits, enabling entangling operations while leaving $A^{\rm r}$ qubits unaffected.
Let us now explicitly write the sequence for the implementations of the Hadamard gate and the C$Z$ gate which the unification of both guarantees universal quantum computation. \\

\paragraph{Hadamard gate.} Following the definitions in Eq.~\eqref{eq:theta-vector} and the general form of the unitaries \eqref{operation unitary}, we have that the Hadamard can be decomposed, depending if the ICC is on a $B$- or $C$-type column, as 

\begin{equation}
    \hat{U}_{\rm Hadamard} := \hat{U}^\dagger_H Z^{\rm tot}_{A^{\rm r}} \hat{U}_H
\end{equation}
with $\hat{U}_H := \hat{U}^{(4)}_{B,C} \hat{U}^{(3)}_{B,C}\hat{U}^{(2)}_{B,C}\hat{U}^{(1)}_{B,C}$
where:


\begin{table}[H]
\centering
\begin{tabular}{lll}
\hline
\textbf{Unitary} & \boldmath$\phi_{B,C}^{(i)}$ & \boldmath$\tau_i$ \\
\hline
$\hat{U}^{(1)}_{B,C}$ & $0$ & $\frac{\pi}{4} \, \Omega_{B,C}^{-1}$ \\
$\hat{U}^{(2)}_{B,C}$ & $\frac{\pi}{2}$ & $\frac{\pi}{2} \, \Omega_{B,C}^{-1}$ \\
$\hat{U}^{(3)}_{B,C}$ & $- \frac{\pi}{4}$ & $\frac{3\pi}{4} \, \Omega_{B,C}^{-1}$ \\
$\hat{U}^{(4)}_{B,C}$ & $\frac{\pi}{4}$ & $\frac{3\pi}{2} \, \Omega_{B,C}^{-1}$ \\
\hline
\end{tabular}
\caption{\textbf{Explicit pulses for Hadamard gate.} Parameters inducing unitaries $\hat{U}^{(i)}_{B,C}$ via control phases $\phi_{B,C}^{(i)}$ and evolution times $\tau_i$.}
\label{tab:unitaries}
\end{table}

and $Z^{\rm tot}_{A^{\rm r}} := \hat{U}_A(\theta=2\pi, \phi=0)$.\\

\paragraph{Control-Phase gate.} To implement the CZ gate, once the ICC is in correspondence with a crossed $A$-type qubit connecting computational qubits, we use the transformation $\hat{Z}_{A^{\times}}^{(\rm tot)}$ that can be realized through a five-step sequence $\hat{U}^{(5)}_{A} \hat{U}^{(4)}_{A}\hat{U}^{(3)}_{A} \hat{U}^{(2)}_{A}\hat{U}^{(1)}_{A}$
 where:
 

\begin{table}[H]
\centering
\begin{tabular}{lll}
\hline
\textbf{Unitary} & \boldmath$\phi_A^{(i)}$ & \boldmath$\tau_i$ \\
\hline
$\hat{U}^{(1)}_{A}$ & $\frac{\pi}{2}$ & $\frac{\pi}{4} \, \Omega_A^{-1}$ \\
$\hat{U}^{(2)}_{A}$ & $0$ & $\pi \, \Omega_A^{-1}$ \\
$\hat{U}^{(3)}_{A}$ & $\frac{\pi}{2}$ & $\frac{\pi}{2} \, \Omega_A^{-1}$ \\
$\hat{U}^{(4)}_{A}$ & $0$ & $\pi \, \Omega_A^{-1}$ \\
$\hat{U}^{(5)}_{A}$ & $\frac{\pi}{2}$ & $\frac{\pi}{4} \, \Omega_A^{-1}$ \\
\hline
\end{tabular}
\caption{\textbf{Explicit pulses for CZ gate.} Parameters inducing unitaries $\hat{U}^{(i)}_{A}$ via control phases $\phi_A^{(i)}$ and evolution times $\tau_i$.}
\label{tab:unitaries_A}
\end{table}

\section{Inhomogeneities in superconducting qubits}\label{app:inhomogenjul}

In this section, we extend the fidelity analysis started in Sec.~\ref{sec:disordermain} of the main text for the quantum information flow and the one- and two-qubit gates in the presence of inhomogeneities in the relevant hardware parameters. The numerical parameters considered for this investigation are shown in Tab. ~\ref{tab:paramsitensor}.
\begin{table}[h]
\centering
\begin{tabular}{ll}
\hline
\textbf{Parameter} & \textbf{Value} \\
\hline
Number of computational qubits & $N = 2$ (i.e., two rows) \\
ZZ coupling, Rabi oscillation ratio & $\eta_{\rm BR} = 20$ \\
Rabi oscillation & $\Omega_{A} = \Omega_{B} = \Omega_{C} = 10~\mathrm{MHz}$ \\
ZZ coupling & $\zeta = (2\pi) \times 200~\mathrm{MHz}$ \\
Qubit resonance frequency & $\omega_A = \omega_B = \omega_C = (2\pi) \times 7000~\mathrm{MHz}$ \\
\hline
\end{tabular}
\caption{\textbf{Numerical simulation parameters.} System parameters used in the numerical simulations with the \texttt{ITensor} package.}
\label{tab:paramsitensor}
\end{table}

\label{sec:disorder}
\subsection{Fidelity of information flow}
Each numerical result presented below is obtained using a tensor-networks-based approach. The Julia version of the \texttt{ITensor} package has been
used~\cite{ITensor, ITensor2}. For all cases, a detailed analysis of the numerical cutoffs has been systematically performed to ensure the convergence of the presented results.
These cutoffs comprise the bond dimension associated with the truncated time-evolved matrix product state $\ket{\Psi(t)}$ denoted by $\bar{a}$, and the time-step $\delta t$.
Finally, we stress that the system size, the energy scales, as well as the disorder strengths considered in the numerical simulations are experimentally relevant for superconducting quantum computing platforms.

We provide now additional information regarding the simulation of the digital twin associated to the ladder architecture as well as the tensor network numerical approach. 
The 2D ladder quantum system interacting through nearest-neighbor antiferromagnetic interactions can be simulated as a one-dimensional (1D) quantum lattice model with non-standard, i.e. non-algebraic and non-preserving-translational-invariance, long-range interactions. The latter approach's efficiency is due to the small number of row-couplers. Indeed
for $N$ rows, the architecture requires $N-1$ couplers. Regarding the tensor network approach to perform the numerical simulations, the latter is based on the MPS (MPO) form to represent
many-body quantum states (operators)~\cite{schollwock2005,schollwock2011}. The real-time evolution, to investigate the dynamics of the ladder architecture, is performed using the time dependent variational principle (TDVP)~\cite{haegeman2011} with ancillary Krylov subspace (AKS)~\cite{yang2020}.

We present our numerical simulations regarding the information flow (IF) in the presence of inhomogeneities, i.e. disorder, for the pulsation associated with each superconducting qubit $\omega_i$ and the nearest-neighbor antiferromagnetic interaction $\zeta$. Adding imperfections such as static disorder in the ladder architecture is a fundamental key concern regarding the robustness of the superconducting architecture. The latter finds its origin from fabrication-induced variations in qubit parameters and thus disrupt coherent quantum information transfer. This explains the importance of the numerical calculations presented here where we analyze the dynamics of the disordered
minimal globally-driven superconducting ladder architecture. 

As previously mentioned, the inhomogeneities on (a) the energy spacing $\omega_i$ and (b) the nearest-neighbor antiferromagnetic ZZ coupling strength $\zeta$ are considered.
The latter are characterized by two distinct normal Gaussian distributions centered on their nominal value, e.g. $\mu = \bar{\omega}$ and where their corresponding width, or equivalently
the full width at half maximum (FWHM), is characterized via the standard deviation $\sigma$. The inhomogeneous energy scales $\omega$ and $\zeta$ are defined via the normal distribution $\mathcal{N}(\mu=\bar{\omega}, \sigma=\epsilon\bar{\omega})$ and
$\mathcal{N}(\mu=\bar{\zeta}, \sigma=\epsilon\bar{\zeta})$ respectively, where $\epsilon$ represents the disorder percentage.

\begin{figure}[h!]
\centering
\includegraphics[scale = 1]{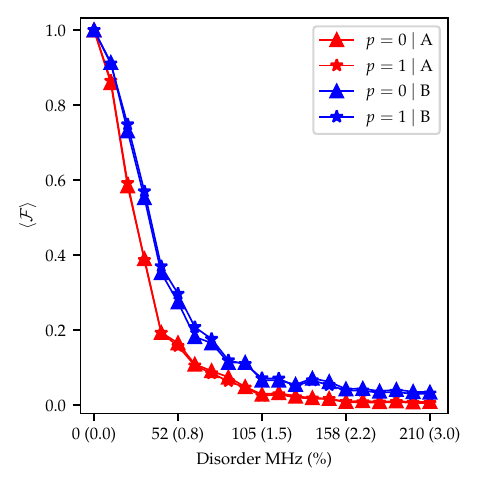}
\vspace{-0.3cm}
\caption{\label{fig:omega}
\textbf{Fidelity w.r.t. disorder percentage:} $\mathcal{F}(p,0) = \vert \langle \Psi_{\mathrm{target}}(p,0) | \Psi_{\mathrm{final}} \rangle \vert$. The RF+RWA drive Hamiltonian for $N = 2$ is considered. 
The initial $A$-type or $B$-type ICC is defined as follows $ \vert \psi_{\mathrm{initial}}(p,\phi) \rangle$. The average over $\omega$-disorder characterized by $\epsilon$ is performed for $N_{\rm samples}= 100$ initial states and $\eta_{\rm BR} = 20$. The independent inhomogeneities are applied to all the superconducting qubits and generated using a gaussian normal distribution $\mathcal{N}(\mu = \bar{\omega}, \sigma = \epsilon \bar{\omega})$. On the $x$-axis is shown the amount of disorder in MHz and in percentage. Here, the $B$-type and $C$-type $\Pi$-pulses, to translate the ICC from one column, are performed simultaneously.}
\end{figure}

\paragraph{Inhomogeneities on the pulsation $\omega$.} We first investigate the dynamics of the disordered $N=2$ globally-driven superconducting quantum architecture. 
For this purpose, the time-dependent fidelity $\mathcal{F}(p,\phi) = \vert \langle \Psi_{\mathrm{target}}(p,\phi) | \Psi_{\rm final} \rangle \vert$ is investigated. The initial ICC is defined as:
\begin{equation}
\vert \psi_{\mathrm{initial}}(p,\phi) \rangle = \big(\sqrt{1-p} \ket{0} + \sqrt{p}e^{i\phi} \ket{1}\big) \otimes  \ket{0}.
\end{equation}
The two local states represent the quantum state associated with each computational qubit forming the ICC. In what follows, we fix $\phi = 0$. The numerical cutoffs are $\bar{a} \propto 10^3$
for the MPS bond dimension, $\lambda = 10^{-20}$ denoting the smallest singular value for the truncation of the MPS and $N_{\chi_{\ell}} = 2$ referring to the number of subdivisions. The latter corresponds to the number of time-steps
to perform in order to numerically simulate each time window $\tau_{\chi_\ell}$ associated with the real-time unitary evolution operator $\hat{U}_{\chi}^{(\ell)}$, i.e.~$N_{\chi_\ell}\delta\mathcal{T}_{\chi_\ell} = \mathcal{T}_{\chi_\ell}$.
Note that to correctly take into account inhomogeneities over the ladder, an average over the disorder needs to be performed; requiring $N_{\rm initial} \in \{100, 200, 300\}$, i.e.~several hundreds, for $\epsilon = 1\%$.  
The approximate RF+RWA Hamiltonian for the global drive is considered for the numerical simulations. 

We first investigate the case of  $\omega$-disorder, i.e., of inhomogeneities on the qubit pulsation $\omega$. 
In Fig.~\ref{fig:omega}, the tensor-network-based numerical calculations show that the fidelity $\mathcal{F}(p, 0)$ does not depend on the scalar $p$. This statement has also been verified for different ICC initial types.
This means that the sampling has been correctly performed and the numerical results have converged well. Note that a similar investigation has been performed for $N_{\rm samples}\in \{50, 100\}$ and similar results
were obtained permitting us to certify the validity of the tensor networks calculations. Surprisingly, the fidelity $\mathcal{F}(p, 0)$ with $p \in \{0,1\}$ significantly differs whether an $A$-type ICC or $B$-type ICC is initially considered.
This can be explained by the sequence length in order to perform the quantum information flow; the latter being more complex for an initial $A$-type ICC compared to an initial $B$-type or $C$-type ICC. Most importantly, we show 
here that the fidelity $\mathcal{F}(p,0)$ decreases when the $\omega$-disorder percentage $\epsilon$ increases. This is consistent with the fact that the more disordered the quantum system is, the less reliable the quantum information flow algorithm is. 
Besides, even at low disorder, the fidelity is very low, i.e. the relative error associated with the fidelity $\mathcal{F}(p, 0)$ is close to $100\%$. This justifies the importance of performing pulse optimization to recover a high fidelity, i.e. $\vert 1 -\mathcal{F}(p, 0)\vert \simeq 10^{-4}$.  

\begin{figure}[h!]
\centering
\includegraphics[scale = 1]{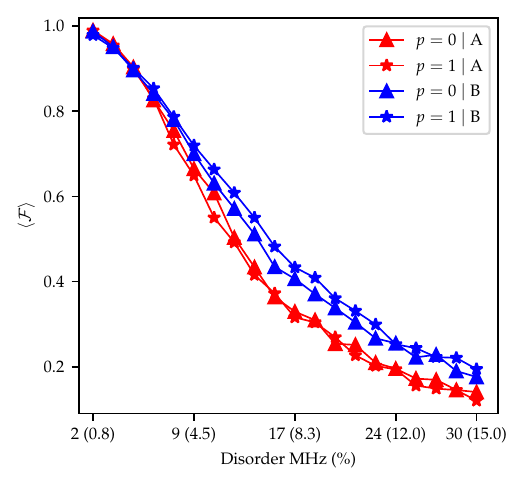}
\vspace{-0.3cm}
\caption{\label{fig:zeta}
\textbf{Fidelity w.r.t. disorder percentage:} $\mathcal{F}(p,0) = \vert \langle \Psi_{\mathrm{target}}(p,0) | \Psi_{\mathrm{final}} \rangle \vert$. The RF+RWA drive Hamiltonian for $N = 2$ is considered. 
The initial $A$-type or $B$-type ICC is defined as follows $ \vert \psi_{\mathrm{initial}}(p,\phi) \rangle$. The average over $\zeta$-disorder characterized by $\epsilon$ is performed for $N_{\rm samples}= 200$ initial states and $\eta_{\rm BR} = 20$. The independent inhomogeneities are applied to all the superconducting qubits and generated using a gaussian normal distribution $\mathcal{N}(\mu = \bar{\zeta}, \sigma = \epsilon \bar{\zeta})$. On the $x$-axis is shown the amount of disorder in MHz and in percentage. Here, the $B$-type and $C$-type pulses, to translate the ICC from one column, are performed simultaneously.}
\end{figure}

\paragraph{Inhomogeneities on the antiferromagnetic coupling $\zeta$.} We now move on to a similar investigation in which a $\zeta$-disorder is considered.  
In Fig.~\ref{fig:zeta}, the fidelity $\mathcal{F}(p, 0)$ is plotted as a function of the $\zeta$-disorder percentage $\epsilon$. As previously, we have verified that the latter quantity does not depend on the scalar $p$ characterizing the initial many-body
quantum state $\ket{\psi_{\rm initial}(p,0)}$ while considering an initial ICC of $A$- or $B$-type. The convergence of the numerical simulations has been also certified by computing $\mathcal{F}(p, 0)$ for different numbers of samples,
i.e. initial states. The latter, being dependent on $\epsilon_{\rm max} = \mathrm{max}(\epsilon)$, permits to take into account the stochasticity introduced by considering inhomogeneities on the nearest-neighbor antiferromagnetic coupling $\zeta$. 
As expected, the fidelity $\mathcal{F}(p,0)$ decreases when the $\zeta$-disorder percentage $\epsilon$ increases.  
Besides, even at low disorder, the fidelity is very low, i.e. the relative error, being equivalent here to the absolute error, associated to the fidelity $\mathcal{F}(p, 0)$ is close to $100\%$. This justifies the importance of performing pulse optimization to recover a high fidelity, i.e. $\vert 1 - \mathcal{F}(p, 0)\vert \simeq 10^{-4}$.  Note, however, that the fidelity decay is much higher for $\omega$-disorder compared to $\zeta$-disorder as shown on Fig.~\ref{fig:omega} and \ref{fig:zeta}. In other words, inhomogeneities on the local pulsation $\omega$ are more binding than those on the antiferromagnetic interaction $\zeta$ due to the involved energy scales. Indeed, since $\omega \gg \zeta$, the short-time dynamics of the disordered globally-driven 
2D ladder architecture is mainly driven by $\omega$-dependent terms, i.e. the local energy of the superconducting qubits.

\paragraph{Inhomogeneities on the coupling $\zeta$ and the pulsation $\omega$.} 
We finally turn to the case where inhomogeneities on all the relevant energy scales are taken into account, i.e. the local energy $\omega$ and the Néel interaction $\zeta$ (up to the Planck constant $\hbar$). As expected from Fig.~\ref{fig:zeta},
we show on Fig.~\ref{fig:both} that $\zeta$-disorder is not essential, i.e. does not play a major role when investigating the short-time dynamics of the architecture, see solid pink and triangle-left blue lines. For this investigation, the convergence of the tensor networks numerical calculations has been verified by considering $2$ distinct values of $N_{\mathrm{states}}$, different initial many-body quantum states through the variable, i.e. scalar, $p$ and for $A$- and $B$-type ICC. 

To sum up, the latter investigation raised a twofold conclusion: (i) the irrelevance of $\zeta$-disorder and (ii) the importance of performing pulse optimization to recover high fidelities for the quantum information flow as well as and quantum algorithms.

\section{Examples of optimal pulses\label{app:pulse example}}
\begin{figure}[h!]
\centering
\includegraphics[scale = 0.9]{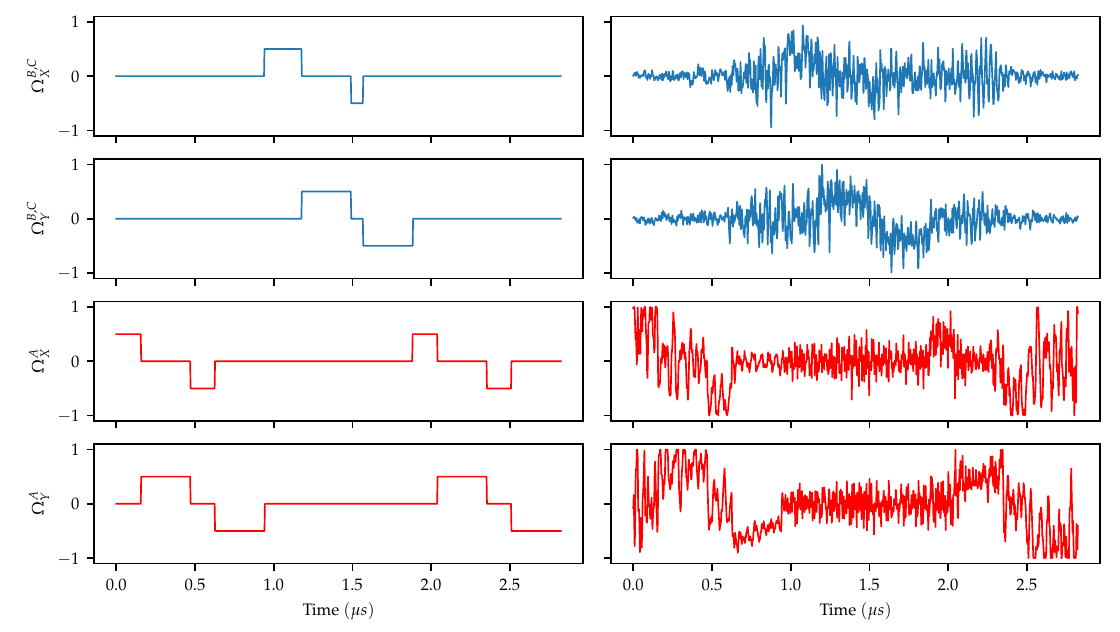}
\vspace{-0.3cm}

\caption{\label{fig:pulse example}
\textbf{Naive and optimal pulse sequence comparison.} This figure shows the naive (left) and optimal (right) pulse sequences for the Information Flow protocol starting from $B$-type qubits. The first two rows ($X$ and $Y$ channel) correspond to the $B$ and $C$ species, while the last two correspond to the $A$ species. It is evident that \texttt{GRAPE} perturbs the initial guess and determines how fluctuations in the naive protocol can be exploited to mitigate the disorder problem. Notably, all species are driven simultaneously, even though the naive protocol drives only one species at a time. }
\end{figure}

We have seen that the \texttt{GRAPE} algorithm is able to find optimal pulse sequences to solve the disorder problem in globally-driven quantum architectures. It is interesting to analyze such optimal sequences to get insight into how it manages to do so. For this purpose, in this section we compare an example of an optimal pulse sequence with the naive protocol, and we show it in Fig.~\ref{fig:pulse example}. The first fact to note is that \texttt{GRAPE} perturbs the naive sequence and exploits small fluctuations to overcome the disorder. Another interesting point is that all species are driven simultaneously in the optimal sequence, even though the naive protocol drives only one species at a time. Moreover, in the naive protocol the absolute value of the amplitude of the channels never surpass $0.5$, while in the optimized one \texttt{GRAPE} makes use of all the space allowed, between $-1$ and $1$.

We conclude that, compared to the non-disordered case where the pulse sequences were of the ``bang-bang'' type, in the disordered case, the optimal global pulses are applied in adjacent time windows but, as we can see in Fig.~\ref{fig:omega}, still keeping their shape.

\clearpage
\section{MPS-based pulse optimization for extended systems}\label{app:mps_grape}

The pulse optimization simulations presented in Sec.~\ref{sec:pulse-optimization-main} were performed on subsystems of 7 qubits using exact state vector methods to maintain computational tractability. However, scaling pulse optimization to larger quantum processors requires more efficient numerical approaches. Matrix product state (MPS) representations offer a natural solution by exploiting the favorable entanglement structure of the ladder architecture, enabling optimization on systems with significantly more qubits while maintaining high accuracy.

We have developed a custom pulse optimization engine that implements the \texttt{GRAPE} methodology~\cite{Khaneja2005_grape} using MPS-based time evolution combined with \texttt{Adam} optimization~\cite{kingma2014adam}. Gradients are computed via forward and backward propagation of quantum states using the time-dependent variational principle (TDVP)~\cite{haegeman2011}.

A key difference in this implementation is the state-based optimization rather than the gate-based optimization. Instead of targeting a specific unitary operator $\hat{U}_{\rm target}$, we specify a training set of initial states $\{|\Psi_{\mathrm{initial}}^{(j)}\rangle\}$ and their corresponding target states $\{|\Psi_{\mathrm{target}}^{(j)}\rangle\}$, then we maximize the average fidelity:
\begin{equation}
\overline{\mathcal{F}} = \frac{1}{N_\mathrm{train}} \sum_{j=1}^{N_\mathrm{train}} 
\big|\langle \Psi_{\mathrm{target}}^{(j)} | \Psi_{\mathrm{final}}^{(j)} \rangle \big|.
\label{eq:mps_cost}
\end{equation}
The control amplitudes are limited to $[-1,1]$ and discretized with time steps on the order of nanoseconds, compatible with the state-of-the-art experimental hardware capabilities~\cite{zurich_instr}.

To demonstrate the capability of this approach, we performed pulse optimization for a Hadamard gate on the full $N=2$ ladder architecture containing 15 physical qubits. The system includes a $2\%$ disorder (single specific realization) in qubit frequencies, with parameters $\bar{\omega} = 2\pi \times 7000$~ MHz, $\bar{\zeta} = 2\pi \times 200$~ MHz, $\Omega_{\chi} = 10$~ MHz and $\eta_{\rm BR} = 20$. The training set consisted of the initial states $|\psi_{\mathrm{initial}}(p,0)\rangle = (\sqrt{1-p}|0\rangle + \sqrt{p}|1\rangle) \otimes |0\rangle$ with $p \in \{0, 0.33, 0.66\}$.

\begin{figure}[h!]
\centering
\includegraphics[width=0.6\columnwidth]{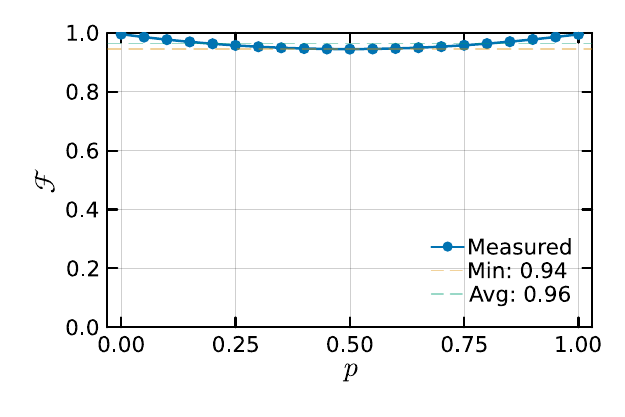}
\caption{\label{fig:hadamard_optimal_mps}
\textbf{MPS-optimized Hadamard gate on $N=2$ ladder.} State fidelity $\mathcal{F}(p,0)$ as a function of initial amplitude $p$ using the pulse sequence optimized by the MPS-\texttt{GRAPE} engine on the full 15-qubit system with $2\%$ static frequency disorder. The optimized sequence achieves average fidelity $\overline{\mathcal{F}} = 0.96$ with worst-case fidelity of $0.94$ in a gate time of $160$~ns, demonstrating both high fidelity and fast operation despite the presence of disorder.}
\end{figure}

\begin{figure}[h!]
\centering
\includegraphics[width=0.6\columnwidth]{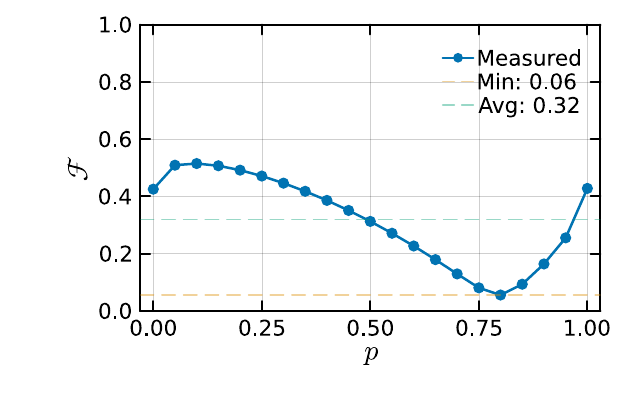}
\caption{\label{fig:hadamard_standard_mps}
\textbf{Standard Hadamard gate on $N=2$ ladder with disorder.} State fidelity $\mathcal{F}(p,0)$ as a function of initial amplitude $p$ using the naive rectangular-pulse protocol on the full 15-qubit system with $2\%$ static frequency disorder. Without pulse optimization, the standard sequence achieves only average fidelity $\overline{\mathcal{F}} = 0.32$ with worst-case fidelity of $0.06$ in $2500$~ns, illustrating the catastrophic effect of disorder on unoptimized global control sequences.}
\end{figure}

Figure~\ref{fig:hadamard_optimal_mps} shows the results for the MPS-optimized pulse sequence, which achieves an average fidelity of $\overline{\mathcal{F}} = 0.96$ with a worst-case fidelity of $0.94$ in all initial states tested. Remarkably, the optimized sequence accomplishes this in just $160$~ns, demonstrating both high fidelity and fast operation despite the presence of $2\%$ static frequency disorder. In contrast, Fig.~\ref{fig:hadamard_standard_mps} shows the performance of the standard protocol under identical disorder conditions, which achieves $\overline{\mathcal{F}} = 0.32$ with a worst-case fidelity of $0.06$ in $2500$~ns. The comparison demonstrates that pulse optimization can simultaneously improve both fidelity and gate speed.

The optimization converged after around 600 iterations, requiring approximately 6 hours of computation time on a standard workstation with 8 CPU cores. This approach is potentially viable for optimizing operations on even larger ladder instances. The successful optimization of the Hadamard gate on the full 15-qubit system provides strong evidence that MPS-based pulse optimization can be extended to practical quantum processors.

These results establish that tensor network methods enable pulse optimization for globally-driven architectures at scales relevant for near-term quantum computing applications. Future work will focus on extending this framework to optimize simultaneous multi-gate operations and investigating the scaling behavior for systems with $N > 1$ computational qubits.


\begin{thebibliography}{51}%
\makeatletter
\providecommand \@ifxundefined [1]{%
 \@ifx{#1\undefined}
}%
\providecommand \@ifnum [1]{%
 \ifnum #1\expandafter \@firstoftwo
 \else \expandafter \@secondoftwo
 \fi
}%
\providecommand \@ifx [1]{%
 \ifx #1\expandafter \@firstoftwo
 \else \expandafter \@secondoftwo
 \fi
}%
\providecommand \natexlab [1]{#1}%
\providecommand \enquote  [1]{``#1''}%
\providecommand \bibnamefont  [1]{#1}%
\providecommand \bibfnamefont [1]{#1}%
\providecommand \citenamefont [1]{#1}%
\providecommand \href@noop [0]{\@secondoftwo}%
\providecommand \href [0]{\begingroup \@sanitize@url \@href}%
\providecommand \@href[1]{\@@startlink{#1}\@@href}%
\providecommand \@@href[1]{\endgroup#1\@@endlink}%
\providecommand \@sanitize@url [0]{\catcode `\\12\catcode `\$12\catcode `\&12\catcode `\#12\catcode `\^12\catcode `\_12\catcode `\%12\relax}%
\providecommand \@@startlink[1]{}%
\providecommand \@@endlink[0]{}%
\providecommand \url  [0]{\begingroup\@sanitize@url \@url }%
\providecommand \@url [1]{\endgroup\@href {#1}{\urlprefix }}%
\providecommand \urlprefix  [0]{URL }%
\providecommand \Eprint [0]{\href }%
\providecommand \doibase [0]{https://doi.org/}%
\providecommand \selectlanguage [0]{\@gobble}%
\providecommand \bibinfo  [0]{\@secondoftwo}%
\providecommand \bibfield  [0]{\@secondoftwo}%
\providecommand \translation [1]{[#1]}%
\providecommand \BibitemOpen [0]{}%
\providecommand \bibitemStop [0]{}%
\providecommand \bibitemNoStop [0]{.\EOS\space}%
\providecommand \EOS [0]{\spacefactor3000\relax}%
\providecommand \BibitemShut  [1]{\csname bibitem#1\endcsname}%
\let\auto@bib@innerbib\@empty
\bibitem [{\citenamefont {Arute}\ and\ \citenamefont {et~al.}(2019)}]{Martinis2019quantum}%
  \BibitemOpen
  \bibfield  {author} {\bibinfo {author} {\bibfnamefont {F.}~\bibnamefont {Arute}}\ and\ \bibinfo {author} {\bibnamefont {et~al.}},\ }\bibfield  {title} {\bibinfo {title} {Quantum supremacy using a programmable superconducting processor},\ }\href {https://doi.org/10.1038/s41586-019-1666-5} {\bibfield  {journal} {\bibinfo  {journal} {Nature}\ }\textbf {\bibinfo {volume} {574}},\ \bibinfo {pages} {505} (\bibinfo {year} {2019})}\BibitemShut {NoStop}%
\bibitem [{\citenamefont {Wu}\ and\ \citenamefont {et~al.}(2021)}]{supremacy_PRL_2021}%
  \BibitemOpen
  \bibfield  {author} {\bibinfo {author} {\bibfnamefont {T.}~\bibnamefont {Wu}}\ and\ \bibinfo {author} {\bibnamefont {et~al.}},\ }\bibfield  {title} {\bibinfo {title} {Strong quantum computational advantage using a superconducting quantum processor},\ }\href {https://doi.org/10.1103/PhysRevLett.127.180501} {\bibfield  {journal} {\bibinfo  {journal} {Phys. Rev. Lett.}\ }\textbf {\bibinfo {volume} {127}},\ \bibinfo {pages} {180501} (\bibinfo {year} {2021})}\BibitemShut {NoStop}%
\bibitem [{\citenamefont {Bravyi}\ \emph {et~al.}(2022)\citenamefont {Bravyi}, \citenamefont {Dial}, \citenamefont {Gambetta}, \citenamefont {Gil},\ and\ \citenamefont {Nazario}}]{bravyi2022future}%
  \BibitemOpen
  \bibfield  {author} {\bibinfo {author} {\bibfnamefont {S.}~\bibnamefont {Bravyi}}, \bibinfo {author} {\bibfnamefont {O.}~\bibnamefont {Dial}}, \bibinfo {author} {\bibfnamefont {J.~M.}\ \bibnamefont {Gambetta}}, \bibinfo {author} {\bibfnamefont {D.}~\bibnamefont {Gil}},\ and\ \bibinfo {author} {\bibfnamefont {Z.}~\bibnamefont {Nazario}},\ }\bibfield  {title} {\bibinfo {title} {The future of quantum computing with superconducting qubits},\ }\href {https://doi.org/10.1063/5.0082975} {\bibfield  {journal} {\bibinfo  {journal} {J. Appl. Phys.}\ }\textbf {\bibinfo {volume} {132}},\ \bibinfo {pages} {160902} (\bibinfo {year} {2022})}\BibitemShut {NoStop}%
\bibitem [{\citenamefont {Ezratty}(2023)}]{ezratty2023perspective}%
  \BibitemOpen
  \bibfield  {author} {\bibinfo {author} {\bibfnamefont {O.}~\bibnamefont {Ezratty}},\ }\bibfield  {title} {\bibinfo {title} {Perspective on superconducting qubit quantum computing},\ }\href {https://doi.org/10.1140/epja/s10050-023-01006-7} {\bibfield  {journal} {\bibinfo  {journal} {Eur. Phys. J. A}\ }\textbf {\bibinfo {volume} {59}},\ \bibinfo {pages} {94} (\bibinfo {year} {2023})}\BibitemShut {NoStop}%
\bibitem [{\citenamefont {Schoelkopf}\ and\ \citenamefont {Girvin}(2008)}]{Girvin2008}%
  \BibitemOpen
  \bibfield  {author} {\bibinfo {author} {\bibfnamefont {R.~J.}\ \bibnamefont {Schoelkopf}}\ and\ \bibinfo {author} {\bibfnamefont {S.~M.}\ \bibnamefont {Girvin}},\ }\bibfield  {title} {\bibinfo {title} {Wiring up quantum systems},\ }\href {https://doi.org/10.1038/451664a} {\bibfield  {journal} {\bibinfo  {journal} {Nature}\ }\textbf {\bibinfo {volume} {451}},\ \bibinfo {pages} {664–669} (\bibinfo {year} {2008})}\BibitemShut {NoStop}%
\bibitem [{\citenamefont {Gambetta}\ \emph {et~al.}(2017)\citenamefont {Gambetta}, \citenamefont {Chow},\ and\ \citenamefont {Steffen}}]{Gambetta2017}%
  \BibitemOpen
  \bibfield  {author} {\bibinfo {author} {\bibfnamefont {J.~M.}\ \bibnamefont {Gambetta}}, \bibinfo {author} {\bibfnamefont {J.~M.}\ \bibnamefont {Chow}},\ and\ \bibinfo {author} {\bibfnamefont {M.}~\bibnamefont {Steffen}},\ }\bibfield  {title} {\bibinfo {title} {Building logical qubits in a superconducting quantum computing system},\ }\href {https://doi.org/10.1038/s41534-016-0004-0} {\bibfield  {journal} {\bibinfo  {journal} {npj Quantum Inf.}\ }\textbf {\bibinfo {volume} {3}},\ \bibinfo {pages} {2} (\bibinfo {year} {2017})}\BibitemShut {NoStop}%
\bibitem [{\citenamefont {Kjaergaard}\ \emph {et~al.}(2020)\citenamefont {Kjaergaard}, \citenamefont {Schwartz}, \citenamefont {Braumüller}, \citenamefont {Krantz}, \citenamefont {Wang}, \citenamefont {Gustavsson},\ and\ \citenamefont {Oliver}}]{Kjaergaard2020}%
  \BibitemOpen
  \bibfield  {author} {\bibinfo {author} {\bibfnamefont {M.}~\bibnamefont {Kjaergaard}}, \bibinfo {author} {\bibfnamefont {M.~E.}\ \bibnamefont {Schwartz}}, \bibinfo {author} {\bibfnamefont {J.}~\bibnamefont {Braumüller}}, \bibinfo {author} {\bibfnamefont {P.}~\bibnamefont {Krantz}}, \bibinfo {author} {\bibfnamefont {J.~I.}\ \bibnamefont {Wang}}, \bibinfo {author} {\bibfnamefont {S.}~\bibnamefont {Gustavsson}},\ and\ \bibinfo {author} {\bibfnamefont {W.~D.}\ \bibnamefont {Oliver}},\ }\bibfield  {title} {\bibinfo {title} {Superconducting qubits: current state of play},\ }\href {https://doi.org/10.1146/annurev-conmatphys-031119-050605} {\bibfield  {journal} {\bibinfo  {journal} {Annu. Rev. Condens. Matter Phys.}\ }\textbf {\bibinfo {volume} {11}},\ \bibinfo {pages} {369} (\bibinfo {year} {2020})}\BibitemShut {NoStop}%
\bibitem [{\citenamefont {Mohseni}\ and\ \citenamefont {et~al.}(2024)}]{Mohseni2024}%
  \BibitemOpen
  \bibfield  {author} {\bibinfo {author} {\bibfnamefont {M.}~\bibnamefont {Mohseni}}\ and\ \bibinfo {author} {\bibnamefont {et~al.}},\ }\bibfield  {title} {\bibinfo {title} {How to build a quantum supercomputer: Scaling challenges and opportunities},\ }\href {https://doi.org/10.48550/arXiv.2411.10406} {\bibfield  {journal} {\bibinfo  {journal} {arXiv:2411.10406}\ } (\bibinfo {year} {2024})}\BibitemShut {NoStop}%
\bibitem [{\citenamefont {Lloyd}(1993)}]{Lloyd_1993}%
  \BibitemOpen
  \bibfield  {author} {\bibinfo {author} {\bibfnamefont {S.}~\bibnamefont {Lloyd}},\ }\bibfield  {title} {\bibinfo {title} {A potentially realizable quantum computer},\ }\href {https://doi.org/10.1126/science.261.5128.1569} {\bibfield  {journal} {\bibinfo  {journal} {Science}\ }\textbf {\bibinfo {volume} {261}},\ \bibinfo {pages} {1569} (\bibinfo {year} {1993})}\BibitemShut {NoStop}%
\bibitem [{\citenamefont {Lloyd}(1999)}]{Lloyd_1993_SI}%
  \BibitemOpen
  \bibfield  {author} {\bibinfo {author} {\bibfnamefont {S.}~\bibnamefont {Lloyd}},\ }\bibfield  {title} {\bibinfo {title} {Programming pulse driven quantum computers},\ }\href {https://doi.org/10.48550/arXiv.quant-ph/9912086} {\bibfield  {journal} {\bibinfo  {journal} {arXiv:quant-ph/9912086}\ } (\bibinfo {year} {1999})}\BibitemShut {NoStop}%
\bibitem [{\citenamefont {Benjamin}(2001{\natexlab{a}})}]{benjamin_2001_1}%
  \BibitemOpen
  \bibfield  {author} {\bibinfo {author} {\bibfnamefont {S.~C.}\ \bibnamefont {Benjamin}},\ }\bibfield  {title} {\bibinfo {title} {Simple pulses for universal quantum computation with a heisenberg abab chain},\ }\href {https://doi.org/10.1103/PhysRevA.64.054303} {\bibfield  {journal} {\bibinfo  {journal} {Phys. Rev. A}\ }\textbf {\bibinfo {volume} {64}},\ \bibinfo {pages} {054303} (\bibinfo {year} {2001}{\natexlab{a}})}\BibitemShut {NoStop}%
\bibitem [{\citenamefont {Benjamin}(2001{\natexlab{b}})}]{benjamin_2001_2}%
  \BibitemOpen
  \bibfield  {author} {\bibinfo {author} {\bibfnamefont {S.~C.}\ \bibnamefont {Benjamin}},\ }\bibfield  {title} {\bibinfo {title} {Quantum computing without local control of qubit-qubit interactions},\ }\href {https://doi.org/10.1103/PhysRevLett.88.017904} {\bibfield  {journal} {\bibinfo  {journal} {Phys. Rev. Lett.}\ }\textbf {\bibinfo {volume} {88}},\ \bibinfo {pages} {017904} (\bibinfo {year} {2001}{\natexlab{b}})}\BibitemShut {NoStop}%
\bibitem [{\citenamefont {Levy}(2002)}]{Levy_2002}%
  \BibitemOpen
  \bibfield  {author} {\bibinfo {author} {\bibfnamefont {J.}~\bibnamefont {Levy}},\ }\bibfield  {title} {\bibinfo {title} {Universal quantum computation with spin-1/2 pairs and heisenberg exchange},\ }\href {https://doi.org/10.1103/PhysRevLett.89.147902} {\bibfield  {journal} {\bibinfo  {journal} {Phys. Rev. Lett.}\ }\textbf {\bibinfo {volume} {89}},\ \bibinfo {pages} {147902} (\bibinfo {year} {2002})}\BibitemShut {NoStop}%
\bibitem [{\citenamefont {Benjamin}\ and\ \citenamefont {Bose}(2003)}]{benjamin_2003}%
  \BibitemOpen
  \bibfield  {author} {\bibinfo {author} {\bibfnamefont {S.~C.}\ \bibnamefont {Benjamin}}\ and\ \bibinfo {author} {\bibfnamefont {S.}~\bibnamefont {Bose}},\ }\bibfield  {title} {\bibinfo {title} {Quantum computing with an always-on heisenberg interaction},\ }\href {https://doi.org/10.1103/PhysRevLett.90.247901} {\bibfield  {journal} {\bibinfo  {journal} {Phys. Rev. Lett.}\ }\textbf {\bibinfo {volume} {90}},\ \bibinfo {pages} {247901} (\bibinfo {year} {2003})}\BibitemShut {NoStop}%
\bibitem [{\citenamefont {Benjamin}(2004)}]{benjamin_2004}%
  \BibitemOpen
  \bibfield  {author} {\bibinfo {author} {\bibfnamefont {S.~C.}\ \bibnamefont {Benjamin}},\ }\bibfield  {title} {\bibinfo {title} {Multi-qubit gates in arrays coupled by 'always-on' interactions},\ }\href {https://doi.org/10.1088/1367-2630/6/1/061} {\bibfield  {journal} {\bibinfo  {journal} {New J. Phys.}\ }\textbf {\bibinfo {volume} {6}},\ \bibinfo {pages} {61} (\bibinfo {year} {2004})}\BibitemShut {NoStop}%
\bibitem [{\citenamefont {Benjamin}\ and\ \citenamefont {Bose}(2004)}]{benjamin-bose_2004}%
  \BibitemOpen
  \bibfield  {author} {\bibinfo {author} {\bibfnamefont {S.~C.}\ \bibnamefont {Benjamin}}\ and\ \bibinfo {author} {\bibfnamefont {S.}~\bibnamefont {Bose}},\ }\bibfield  {title} {\bibinfo {title} {Quantum computing in arrays coupled by “always-on” interactions},\ }\href {https://doi.org/10.1103/PhysRevA.70.032314} {\bibfield  {journal} {\bibinfo  {journal} {Phys. Rev. A}\ }\textbf {\bibinfo {volume} {70}},\ \bibinfo {pages} {032314} (\bibinfo {year} {2004})}\BibitemShut {NoStop}%
\bibitem [{\citenamefont {Ivanyos}\ \emph {et~al.}(2005)\citenamefont {Ivanyos}, \citenamefont {Massar},\ and\ \citenamefont {Nagy}}]{Ivanyos_2005}%
  \BibitemOpen
  \bibfield  {author} {\bibinfo {author} {\bibfnamefont {G.}~\bibnamefont {Ivanyos}}, \bibinfo {author} {\bibfnamefont {S.}~\bibnamefont {Massar}},\ and\ \bibinfo {author} {\bibfnamefont {A.~B.}\ \bibnamefont {Nagy}},\ }\bibfield  {title} {\bibinfo {title} {Quantum computing on lattices using global two-qubit gates},\ }\href {https://doi.org/10.1103/PhysRevA.72.022339} {\bibfield  {journal} {\bibinfo  {journal} {Phys. Rev. A}\ }\textbf {\bibinfo {volume} {72}},\ \bibinfo {pages} {022339} (\bibinfo {year} {2005})}\BibitemShut {NoStop}%
\bibitem [{\citenamefont {Kay}\ and\ \citenamefont {Pachos}(2004)}]{Kay_2004}%
  \BibitemOpen
  \bibfield  {author} {\bibinfo {author} {\bibfnamefont {A.}~\bibnamefont {Kay}}\ and\ \bibinfo {author} {\bibfnamefont {J.~K.}\ \bibnamefont {Pachos}},\ }\bibfield  {title} {\bibinfo {title} {Quantum computation in optical lattices via global laser addressing},\ }\href {https://doi.org/10.1088/1367-2630/6/1/126} {\bibfield  {journal} {\bibinfo  {journal} {New J. Phys.}\ }\textbf {\bibinfo {volume} {6}},\ \bibinfo {pages} {126} (\bibinfo {year} {2004})}\BibitemShut {NoStop}%
\bibitem [{\citenamefont {Fitzsimons}\ and\ \citenamefont {Twamley}(2006)}]{Fitzsimons_2006}%
  \BibitemOpen
  \bibfield  {author} {\bibinfo {author} {\bibfnamefont {J.}~\bibnamefont {Fitzsimons}}\ and\ \bibinfo {author} {\bibfnamefont {J.}~\bibnamefont {Twamley}},\ }\bibfield  {title} {\bibinfo {title} {Globally controlled quantum wires for perfect qubit transport, mirroring, and computing},\ }\href {https://doi.org/10.1103/PhysRevLett.97.090502} {\bibfield  {journal} {\bibinfo  {journal} {Phys. Rev. Lett.}\ }\textbf {\bibinfo {volume} {97}},\ \bibinfo {pages} {090502} (\bibinfo {year} {2006})}\BibitemShut {NoStop}%
\bibitem [{\citenamefont {Paz-Silva}\ \emph {et~al.}(2009)\citenamefont {Paz-Silva}, \citenamefont {Brennen},\ and\ \citenamefont {Twamley}}]{Silva_2009}%
  \BibitemOpen
  \bibfield  {author} {\bibinfo {author} {\bibfnamefont {G.~A.}\ \bibnamefont {Paz-Silva}}, \bibinfo {author} {\bibfnamefont {G.~K.}\ \bibnamefont {Brennen}},\ and\ \bibinfo {author} {\bibfnamefont {J.}~\bibnamefont {Twamley}},\ }\bibfield  {title} {\bibinfo {title} {Globally controlled universal quantum computation with arbitrary subsystem dimension},\ }\href {https://doi.org/10.1103/PhysRevA.80.052318} {\bibfield  {journal} {\bibinfo  {journal} {Phys. Rev. A}\ }\textbf {\bibinfo {volume} {80}},\ \bibinfo {pages} {052318} (\bibinfo {year} {2009})}\BibitemShut {NoStop}%
\bibitem [{\citenamefont {Viola}(2002)}]{viola}%
  \BibitemOpen
  \bibfield  {author} {\bibinfo {author} {\bibfnamefont {L.}~\bibnamefont {Viola}},\ }\bibfield  {title} {\bibinfo {title} {Quantum control via encoded dynamical decoupling},\ }\href {http://dx.doi.org/10.1103/PhysRevA.66.012307} {\bibfield  {journal} {\bibinfo  {journal} {Physical Review A}\ }\textbf {\bibinfo {volume} {66}} (\bibinfo {year} {2002})}\BibitemShut {NoStop}%
\bibitem [{\citenamefont {Cesa}\ and\ \citenamefont {Pichler}(2023)}]{cesa2023universal}%
  \BibitemOpen
  \bibfield  {author} {\bibinfo {author} {\bibfnamefont {F.}~\bibnamefont {Cesa}}\ and\ \bibinfo {author} {\bibfnamefont {H.}~\bibnamefont {Pichler}},\ }\bibfield  {title} {\bibinfo {title} {Universal quantum computation in globally driven rydberg atom arrays},\ }\href {https://doi.org/10.1103/PhysRevLett.131.170601} {\bibfield  {journal} {\bibinfo  {journal} {Phys. Rev. Lett.}\ }\textbf {\bibinfo {volume} {131}},\ \bibinfo {pages} {170691} (\bibinfo {year} {2023})}\BibitemShut {NoStop}%
\bibitem [{\citenamefont {Patomaki}\ \emph {et~al.}(2024)\citenamefont {Patomaki}, \citenamefont {Gonzalez-Zalba}, \citenamefont {Fogarty}, \citenamefont {Cai}, \citenamefont {Benjamin},\ and\ \citenamefont {Morton}}]{Patomaki_2024}%
  \BibitemOpen
  \bibfield  {author} {\bibinfo {author} {\bibfnamefont {S.~M.}\ \bibnamefont {Patomaki}}, \bibinfo {author} {\bibfnamefont {M.~F.}\ \bibnamefont {Gonzalez-Zalba}}, \bibinfo {author} {\bibfnamefont {M.~A.}\ \bibnamefont {Fogarty}}, \bibinfo {author} {\bibfnamefont {Z.}~\bibnamefont {Cai}}, \bibinfo {author} {\bibfnamefont {S.~C.}\ \bibnamefont {Benjamin}},\ and\ \bibinfo {author} {\bibfnamefont {J.~J.~L.}\ \bibnamefont {Morton}},\ }\bibfield  {title} {\bibinfo {title} {Pipeline quantum processor architecture for silicon spin qubits},\ }\href {https://doi.org/10.1038/s41534-024-00823-y} {\bibfield  {journal} {\bibinfo  {journal} {npj Quantum Inf.}\ }\textbf {\bibinfo {volume} {10}},\ \bibinfo {pages} {31} (\bibinfo {year} {2024})}\BibitemShut {NoStop}%
\bibitem [{\citenamefont {Menta}\ \emph {et~al.}(2025)\citenamefont {Menta}, \citenamefont {Cioni}, \citenamefont {Aiudi}, \citenamefont {Polini},\ and\ \citenamefont {Giovannetti}}]{menta2024globally}%
  \BibitemOpen
  \bibfield  {author} {\bibinfo {author} {\bibfnamefont {R.}~\bibnamefont {Menta}}, \bibinfo {author} {\bibfnamefont {F.}~\bibnamefont {Cioni}}, \bibinfo {author} {\bibfnamefont {R.}~\bibnamefont {Aiudi}}, \bibinfo {author} {\bibfnamefont {M.}~\bibnamefont {Polini}},\ and\ \bibinfo {author} {\bibfnamefont {V.}~\bibnamefont {Giovannetti}},\ }\bibfield  {title} {\bibinfo {title} {Globally driven superconducting quantum computing architecture},\ }\href {https://doi.org/10.1103/PhysRevResearch.7.L012065} {\bibfield  {journal} {\bibinfo  {journal} {Phys. Rev. Res.}\ }\textbf {\bibinfo {volume} {7}},\ \bibinfo {pages} {L012065} (\bibinfo {year} {2025})}\BibitemShut {NoStop}%
\bibitem [{\citenamefont {Cioni}\ \emph {et~al.}(2024)\citenamefont {Cioni}, \citenamefont {Menta}, \citenamefont {Aiudi}, \citenamefont {Polini},\ and\ \citenamefont {Giovannetti}}]{cioni2024conveyorbelt}%
  \BibitemOpen
  \bibfield  {author} {\bibinfo {author} {\bibfnamefont {F.}~\bibnamefont {Cioni}}, \bibinfo {author} {\bibfnamefont {R.}~\bibnamefont {Menta}}, \bibinfo {author} {\bibfnamefont {R.}~\bibnamefont {Aiudi}}, \bibinfo {author} {\bibfnamefont {M.}~\bibnamefont {Polini}},\ and\ \bibinfo {author} {\bibfnamefont {V.}~\bibnamefont {Giovannetti}},\ }\bibfield  {title} {\bibinfo {title} {Conveyor belt superconducting quantum computer},\ }\href {https://doi.org/10.48550/arXiv.2412.11782} {\bibfield  {journal} {\bibinfo  {journal} {arXiv:2412.11782}\ } (\bibinfo {year} {2024})}\BibitemShut {NoStop}%
\bibitem [{\citenamefont {Menta}\ \emph {et~al.}(2026)\citenamefont {Menta}, \citenamefont {Cioni}, \citenamefont {Aiudi}, \citenamefont {Caravelli}, \citenamefont {Polini},\ and\ \citenamefont {Giovannetti}}]{menta2025building}%
  \BibitemOpen
  \bibfield  {author} {\bibinfo {author} {\bibfnamefont {R.}~\bibnamefont {Menta}}, \bibinfo {author} {\bibfnamefont {F.}~\bibnamefont {Cioni}}, \bibinfo {author} {\bibfnamefont {R.}~\bibnamefont {Aiudi}}, \bibinfo {author} {\bibfnamefont {F.}~\bibnamefont {Caravelli}}, \bibinfo {author} {\bibfnamefont {M.}~\bibnamefont {Polini}},\ and\ \bibinfo {author} {\bibfnamefont {V.}~\bibnamefont {Giovannetti}},\ }\bibfield  {title} {\bibinfo {title} {Building globally controlled quantum processors with $zz$ interactions},\ }\href {https://doi.org/10.1103/lz5d-lnz2} {\bibfield  {journal} {\bibinfo  {journal} {Phys. Rev. A}\ }\textbf {\bibinfo {volume} {113}},\ \bibinfo {pages} {012614} (\bibinfo {year} {2026})}\BibitemShut {NoStop}%
\bibitem [{\citenamefont {Nielsen}\ and\ \citenamefont {Chuang}(2010)}]{Nielsen2010}%
  \BibitemOpen
  \bibfield  {author} {\bibinfo {author} {\bibfnamefont {M.~A.}\ \bibnamefont {Nielsen}}\ and\ \bibinfo {author} {\bibfnamefont {I.~L.}\ \bibnamefont {Chuang}},\ }\bibfield  {title} {\bibinfo {title} {Quantum computation and quantum information},\ }\href@noop {} {\bibfield  {journal} {\bibinfo  {journal} {Cambridge University Press}\ } (\bibinfo {year} {2010})}\BibitemShut {NoStop}%
\bibitem [{\citenamefont {Fishman}\ \emph {et~al.}(2022{\natexlab{a}})\citenamefont {Fishman}, \citenamefont {White},\ and\ \citenamefont {Stoudenmire}}]{ITensor}%
  \BibitemOpen
  \bibfield  {author} {\bibinfo {author} {\bibfnamefont {M.}~\bibnamefont {Fishman}}, \bibinfo {author} {\bibfnamefont {S.~R.}\ \bibnamefont {White}},\ and\ \bibinfo {author} {\bibfnamefont {E.~M.}\ \bibnamefont {Stoudenmire}},\ }\bibfield  {title} {\bibinfo {title} {{The ITensor Software Library for Tensor Network Calculations}},\ }\href {https://doi.org/10.21468/SciPostPhysCodeb.4} {\bibfield  {journal} {\bibinfo  {journal} {SciPost Phys. Codebases}\ ,\ \bibinfo {pages} {4}} (\bibinfo {year} {2022}{\natexlab{a}})}\BibitemShut {NoStop}%
\bibitem [{\citenamefont {Fishman}\ \emph {et~al.}(2022{\natexlab{b}})\citenamefont {Fishman}, \citenamefont {White},\ and\ \citenamefont {Stoudenmire}}]{ITensor2}%
  \BibitemOpen
  \bibfield  {author} {\bibinfo {author} {\bibfnamefont {M.}~\bibnamefont {Fishman}}, \bibinfo {author} {\bibfnamefont {S.~R.}\ \bibnamefont {White}},\ and\ \bibinfo {author} {\bibfnamefont {E.~M.}\ \bibnamefont {Stoudenmire}},\ }\bibfield  {title} {\bibinfo {title} {{Codebase release 0.3 for ITensor}},\ }\href {https://doi.org/10.21468/SciPostPhysCodeb.4-r0.3} {\bibfield  {journal} {\bibinfo  {journal} {SciPost Phys. Codebases}\ ,\ \bibinfo {pages} {4}} (\bibinfo {year} {2022}{\natexlab{b}})}\BibitemShut {NoStop}%
\bibitem [{\citenamefont {Schollw\"ock}(2005)}]{schollwock2005}%
  \BibitemOpen
  \bibfield  {author} {\bibinfo {author} {\bibfnamefont {U.}~\bibnamefont {Schollw\"ock}},\ }\bibfield  {title} {\bibinfo {title} {The density-matrix renormalization group},\ }\href {https://doi.org/10.1103/RevModPhys.77.259} {\bibfield  {journal} {\bibinfo  {journal} {Rev. Mod. Phys.}\ }\textbf {\bibinfo {volume} {77}},\ \bibinfo {pages} {259} (\bibinfo {year} {2005})}\BibitemShut {NoStop}%
\bibitem [{\citenamefont {Schollwöck}(2011)}]{schollwock2011}%
  \BibitemOpen
  \bibfield  {author} {\bibinfo {author} {\bibfnamefont {U.}~\bibnamefont {Schollwöck}},\ }\bibfield  {title} {\bibinfo {title} {The density-matrix renormalization group in the age of matrix product states},\ }\href {https://doi.org/https://doi.org/10.1016/j.aop.2010.09.012} {\bibfield  {journal} {\bibinfo  {journal} {Annals of Physics}\ }\textbf {\bibinfo {volume} {326}},\ \bibinfo {pages} {96} (\bibinfo {year} {2011})}\BibitemShut {NoStop}%
\bibitem [{\citenamefont {Haegeman}\ \emph {et~al.}(2011)\citenamefont {Haegeman}, \citenamefont {Cirac}, \citenamefont {Osborne}, \citenamefont {Pi\ifmmode~\check{z}\else \v{z}\fi{}orn}, \citenamefont {Verschelde},\ and\ \citenamefont {Verstraete}}]{haegeman2011}%
  \BibitemOpen
  \bibfield  {author} {\bibinfo {author} {\bibfnamefont {J.}~\bibnamefont {Haegeman}}, \bibinfo {author} {\bibfnamefont {J.~I.}\ \bibnamefont {Cirac}}, \bibinfo {author} {\bibfnamefont {T.~J.}\ \bibnamefont {Osborne}}, \bibinfo {author} {\bibfnamefont {I.}~\bibnamefont {Pi\ifmmode~\check{z}\else \v{z}\fi{}orn}}, \bibinfo {author} {\bibfnamefont {H.}~\bibnamefont {Verschelde}},\ and\ \bibinfo {author} {\bibfnamefont {F.}~\bibnamefont {Verstraete}},\ }\bibfield  {title} {\bibinfo {title} {Time-dependent variational principle for quantum lattices},\ }\href {https://doi.org/10.1103/PhysRevLett.107.070601} {\bibfield  {journal} {\bibinfo  {journal} {Phys. Rev. Lett.}\ }\textbf {\bibinfo {volume} {107}},\ \bibinfo {pages} {070601} (\bibinfo {year} {2011})}\BibitemShut {NoStop}%
\bibitem [{\citenamefont {Yang}\ and\ \citenamefont {White}(2020)}]{yang2020}%
  \BibitemOpen
  \bibfield  {author} {\bibinfo {author} {\bibfnamefont {M.}~\bibnamefont {Yang}}\ and\ \bibinfo {author} {\bibfnamefont {S.~R.}\ \bibnamefont {White}},\ }\bibfield  {title} {\bibinfo {title} {Time-dependent variational principle with ancillary krylov subspace},\ }\href {https://doi.org/10.1103/PhysRevB.102.094315} {\bibfield  {journal} {\bibinfo  {journal} {Phys. Rev. B}\ }\textbf {\bibinfo {volume} {102}},\ \bibinfo {pages} {094315} (\bibinfo {year} {2020})}\BibitemShut {NoStop}%
\bibitem [{\citenamefont {Khaneja}\ \emph {et~al.}(2005)\citenamefont {Khaneja}, \citenamefont {Reiss}, \citenamefont {Kehlet}, \citenamefont {Schulte-Herbr\"{u}ggen},\ and\ \citenamefont {Glaser}}]{Khaneja2005_grape}%
  \BibitemOpen
  \bibfield  {author} {\bibinfo {author} {\bibfnamefont {N.}~\bibnamefont {Khaneja}}, \bibinfo {author} {\bibfnamefont {T.}~\bibnamefont {Reiss}}, \bibinfo {author} {\bibfnamefont {C.}~\bibnamefont {Kehlet}}, \bibinfo {author} {\bibfnamefont {T.}~\bibnamefont {Schulte-Herbr\"{u}ggen}},\ and\ \bibinfo {author} {\bibfnamefont {S.~J.}\ \bibnamefont {Glaser}},\ }\bibfield  {title} {\bibinfo {title} {Optimal control of coupled spin dynamics: design of nmr pulse sequences by gradient ascent algorithms},\ }\href {https://doi.org/10.1016/j.jmr.2004.11.004} {\bibfield  {journal} {\bibinfo  {journal} {Journal of Magnetic Resonance}\ }\textbf {\bibinfo {volume} {172}},\ \bibinfo {pages} {296–305} (\bibinfo {year} {2005})}\BibitemShut {NoStop}%
\bibitem [{\citenamefont {Ferrie}(2014)}]{ferrie}%
  \BibitemOpen
  \bibfield  {author} {\bibinfo {author} {\bibfnamefont {C.}~\bibnamefont {Ferrie}},\ }\bibfield  {title} {\bibinfo {title} {Self-guided quantum tomography},\ }\href {https://doi.org/10.1103/PhysRevLett.113.190404} {\bibfield  {journal} {\bibinfo  {journal} {Phys. Rev. Lett.}\ }\textbf {\bibinfo {volume} {113}},\ \bibinfo {pages} {190404} (\bibinfo {year} {2014})}\BibitemShut {NoStop}%
\bibitem [{\citenamefont {Turinici}(2019)}]{turinici}%
  \BibitemOpen
  \bibfield  {author} {\bibinfo {author} {\bibfnamefont {G.}~\bibnamefont {Turinici}},\ }\bibfield  {title} {\bibinfo {title} {Stochastic learning control of inhomogeneous quantum ensembles},\ }\href {https://doi.org/10.1103/PhysRevA.100.053403} {\bibfield  {journal} {\bibinfo  {journal} {Phys. Rev. A}\ }\textbf {\bibinfo {volume} {100}},\ \bibinfo {pages} {053403} (\bibinfo {year} {2019})}\BibitemShut {NoStop}%
\bibitem [{\citenamefont {Krotov}(1993)}]{krotov}%
  \BibitemOpen
  \bibfield  {author} {\bibinfo {author} {\bibfnamefont {V.~F.}\ \bibnamefont {Krotov}},\ }\bibfield  {title} {\bibinfo {title} {Global methods in optimal control theory},\ }\href {https://doi.org/10.1007/978-1-4612-0349-0_3} {\bibfield  {journal} {\bibinfo  {journal} {Advances in Nonlinear Dynamics and Control: A Report from Russia}\ ,\ \bibinfo {pages} {74}} (\bibinfo {year} {1993})}\BibitemShut {NoStop}%
\bibitem [{\citenamefont {Goerz}\ \emph {et~al.}(2019)\citenamefont {Goerz}, \citenamefont {Basilewitsch}, \citenamefont {Gago-Encinas}, \citenamefont {Krauss}, \citenamefont {Horn}, \citenamefont {Reich},\ and\ \citenamefont {Koch}}]{goerz}%
  \BibitemOpen
  \bibfield  {author} {\bibinfo {author} {\bibfnamefont {M.~H.}\ \bibnamefont {Goerz}}, \bibinfo {author} {\bibfnamefont {D.}~\bibnamefont {Basilewitsch}}, \bibinfo {author} {\bibfnamefont {F.}~\bibnamefont {Gago-Encinas}}, \bibinfo {author} {\bibfnamefont {M.~G.}\ \bibnamefont {Krauss}}, \bibinfo {author} {\bibfnamefont {K.~P.}\ \bibnamefont {Horn}}, \bibinfo {author} {\bibfnamefont {D.~M.}\ \bibnamefont {Reich}},\ and\ \bibinfo {author} {\bibfnamefont {C.~P.}\ \bibnamefont {Koch}},\ }\bibfield  {title} {\bibinfo {title} {{Krotov: A Python implementation of Krotov's method for quantum optimal control}},\ }\href {https://doi.org/10.21468/SciPostPhys.7.6.080} {\bibfield  {journal} {\bibinfo  {journal} {SciPost Phys.}\ }\textbf {\bibinfo {volume} {7}},\ \bibinfo {pages} {080} (\bibinfo {year} {2019})}\BibitemShut {NoStop}%
\bibitem [{\citenamefont {August}\ and\ \citenamefont {Hern{\'a}ndez-Lobato}(2018)}]{august}%
  \BibitemOpen
  \bibfield  {author} {\bibinfo {author} {\bibfnamefont {M.}~\bibnamefont {August}}\ and\ \bibinfo {author} {\bibfnamefont {J.~M.}\ \bibnamefont {Hern{\'a}ndez-Lobato}},\ }\bibfield  {title} {\bibinfo {title} {Taking gradients through experiments: Lstms and memory proximal policy optimization for black-box quantum control},\ }\href {https://doi.org/10.1007/978-3-030-02465-9_43} {\bibfield  {journal} {\bibinfo  {journal} {High Performance Computing}\ ,\ \bibinfo {pages} {591}} (\bibinfo {year} {2018})}\BibitemShut {NoStop}%
\bibitem [{\citenamefont {Bukov}\ \emph {et~al.}(2018)\citenamefont {Bukov}, \citenamefont {Day}, \citenamefont {Sels}, \citenamefont {Weinberg}, \citenamefont {Polkovnikov},\ and\ \citenamefont {Mehta}}]{bukov}%
  \BibitemOpen
  \bibfield  {author} {\bibinfo {author} {\bibfnamefont {M.}~\bibnamefont {Bukov}}, \bibinfo {author} {\bibfnamefont {A.~G.~R.}\ \bibnamefont {Day}}, \bibinfo {author} {\bibfnamefont {D.}~\bibnamefont {Sels}}, \bibinfo {author} {\bibfnamefont {P.}~\bibnamefont {Weinberg}}, \bibinfo {author} {\bibfnamefont {A.}~\bibnamefont {Polkovnikov}},\ and\ \bibinfo {author} {\bibfnamefont {P.}~\bibnamefont {Mehta}},\ }\bibfield  {title} {\bibinfo {title} {Reinforcement learning in different phases of quantum control},\ }\href {https://doi.org/10.1103/PhysRevX.8.031086} {\bibfield  {journal} {\bibinfo  {journal} {Phys. Rev. X}\ }\textbf {\bibinfo {volume} {8}},\ \bibinfo {pages} {031086} (\bibinfo {year} {2018})}\BibitemShut {NoStop}%
\bibitem [{\citenamefont {Doria}\ \emph {et~al.}(2011)\citenamefont {Doria}, \citenamefont {Calarco},\ and\ \citenamefont {Montangero}}]{monta1}%
  \BibitemOpen
  \bibfield  {author} {\bibinfo {author} {\bibfnamefont {P.}~\bibnamefont {Doria}}, \bibinfo {author} {\bibfnamefont {T.}~\bibnamefont {Calarco}},\ and\ \bibinfo {author} {\bibfnamefont {S.}~\bibnamefont {Montangero}},\ }\bibfield  {title} {\bibinfo {title} {Optimal control technique for many-body quantum dynamics},\ }\href {https://doi.org/10.1103/PhysRevLett.106.190501} {\bibfield  {journal} {\bibinfo  {journal} {Phys. Rev. Lett.}\ }\textbf {\bibinfo {volume} {106}},\ \bibinfo {pages} {190501} (\bibinfo {year} {2011})}\BibitemShut {NoStop}%
\bibitem [{\citenamefont {Rach}\ \emph {et~al.}(2015)\citenamefont {Rach}, \citenamefont {M\"uller}, \citenamefont {Calarco},\ and\ \citenamefont {Montangero}}]{monta2}%
  \BibitemOpen
  \bibfield  {author} {\bibinfo {author} {\bibfnamefont {N.}~\bibnamefont {Rach}}, \bibinfo {author} {\bibfnamefont {M.~M.}\ \bibnamefont {M\"uller}}, \bibinfo {author} {\bibfnamefont {T.}~\bibnamefont {Calarco}},\ and\ \bibinfo {author} {\bibfnamefont {S.}~\bibnamefont {Montangero}},\ }\bibfield  {title} {\bibinfo {title} {Dressing the chopped-random-basis optimization: A bandwidth-limited access to the trap-free landscape},\ }\href {https://doi.org/10.1103/PhysRevA.92.062343} {\bibfield  {journal} {\bibinfo  {journal} {Phys. Rev. A}\ }\textbf {\bibinfo {volume} {92}},\ \bibinfo {pages} {062343} (\bibinfo {year} {2015})}\BibitemShut {NoStop}%
\bibitem [{\citenamefont {Kelly}\ \emph {et~al.}(2014)\citenamefont {Kelly}, \citenamefont {Barends}, \citenamefont {Campbell}, \citenamefont {Chen}, \citenamefont {Chen}, \citenamefont {Chiaro}, \citenamefont {Dunsworth}, \citenamefont {Fowler}, \citenamefont {Hoi}, \citenamefont {Jeffrey}, \citenamefont {Megrant}, \citenamefont {Mutus}, \citenamefont {Neill}, \citenamefont {O'Malley}, \citenamefont {Quintana}, \citenamefont {Roushan}, \citenamefont {Sank}, \citenamefont {Vainsencher}, \citenamefont {Wenner}, \citenamefont {White}, \citenamefont {Cleland},\ and\ \citenamefont {Martinis}}]{kelly}%
  \BibitemOpen
  \bibfield  {author} {\bibinfo {author} {\bibfnamefont {J.}~\bibnamefont {Kelly}}, \bibinfo {author} {\bibfnamefont {R.}~\bibnamefont {Barends}}, \bibinfo {author} {\bibfnamefont {B.}~\bibnamefont {Campbell}}, \bibinfo {author} {\bibfnamefont {Y.}~\bibnamefont {Chen}}, \bibinfo {author} {\bibfnamefont {Z.}~\bibnamefont {Chen}}, \bibinfo {author} {\bibfnamefont {B.}~\bibnamefont {Chiaro}}, \bibinfo {author} {\bibfnamefont {A.}~\bibnamefont {Dunsworth}}, \bibinfo {author} {\bibfnamefont {A.~G.}\ \bibnamefont {Fowler}}, \bibinfo {author} {\bibfnamefont {I.-C.}\ \bibnamefont {Hoi}}, \bibinfo {author} {\bibfnamefont {E.}~\bibnamefont {Jeffrey}}, \bibinfo {author} {\bibfnamefont {A.}~\bibnamefont {Megrant}}, \bibinfo {author} {\bibfnamefont {J.}~\bibnamefont {Mutus}}, \bibinfo {author} {\bibfnamefont {C.}~\bibnamefont {Neill}}, \bibinfo {author} {\bibfnamefont {P.~J.~J.}\ \bibnamefont {O'Malley}}, \bibinfo {author} {\bibfnamefont {C.}~\bibnamefont {Quintana}}, \bibinfo {author} {\bibfnamefont {P.}~\bibnamefont
  {Roushan}}, \bibinfo {author} {\bibfnamefont {D.}~\bibnamefont {Sank}}, \bibinfo {author} {\bibfnamefont {A.}~\bibnamefont {Vainsencher}}, \bibinfo {author} {\bibfnamefont {J.}~\bibnamefont {Wenner}}, \bibinfo {author} {\bibfnamefont {T.~C.}\ \bibnamefont {White}}, \bibinfo {author} {\bibfnamefont {A.~N.}\ \bibnamefont {Cleland}},\ and\ \bibinfo {author} {\bibfnamefont {J.~M.}\ \bibnamefont {Martinis}},\ }\bibfield  {title} {\bibinfo {title} {Optimal quantum control using randomized benchmarking},\ }\href {https://doi.org/10.1103/PhysRevLett.112.240504} {\bibfield  {journal} {\bibinfo  {journal} {Phys. Rev. Lett.}\ }\textbf {\bibinfo {volume} {112}},\ \bibinfo {pages} {240504} (\bibinfo {year} {2014})}\BibitemShut {NoStop}%
\bibitem [{\citenamefont {Gambetta}\ \emph {et~al.}(2011)\citenamefont {Gambetta}, \citenamefont {Motzoi}, \citenamefont {Merkel},\ and\ \citenamefont {Wilhelm}}]{Gambetta2011_drive}%
  \BibitemOpen
  \bibfield  {author} {\bibinfo {author} {\bibfnamefont {J.~M.}\ \bibnamefont {Gambetta}}, \bibinfo {author} {\bibfnamefont {F.}~\bibnamefont {Motzoi}}, \bibinfo {author} {\bibfnamefont {S.~T.}\ \bibnamefont {Merkel}},\ and\ \bibinfo {author} {\bibfnamefont {F.~K.}\ \bibnamefont {Wilhelm}},\ }\bibfield  {title} {\bibinfo {title} {Analytic control methods for high-fidelity unitary operations in a weakly nonlinear oscillator},\ }\href {https://doi.org/10.1103/PhysRevA.83.012308} {\bibfield  {journal} {\bibinfo  {journal} {Phys. Rev. A}\ }\textbf {\bibinfo {volume} {83}},\ \bibinfo {pages} {012308} (\bibinfo {year} {2011})}\BibitemShut {NoStop}%
\bibitem [{\citenamefont {2025}()}]{zurich_instr}%
  \BibitemOpen
  \bibfield  {author} {\bibinfo {author} {\bibfnamefont {Z.~I.~A.}\ \bibnamefont {2025}},\ }\href {https://www.zhinst.com/europe/en/products/shfqc-qubit-controller#specifications} {\bibinfo {title} {Zurich instruments specifications}}\BibitemShut {NoStop}%
\bibitem [{\citenamefont {Lambert}\ \emph {et~al.}(2024)\citenamefont {Lambert}, \citenamefont {Gigu{\`e}re}, \citenamefont {Menczel}, \citenamefont {Li}, \citenamefont {Hopf}, \citenamefont {Su{\'a}rez}, \citenamefont {Gali}, \citenamefont {Lishman}, \citenamefont {Gadhvi}, \citenamefont {Agarwal} \emph {et~al.}}]{lambert2024qutip5quantumtoolbox}%
  \BibitemOpen
  \bibfield  {author} {\bibinfo {author} {\bibfnamefont {N.}~\bibnamefont {Lambert}}, \bibinfo {author} {\bibfnamefont {E.}~\bibnamefont {Gigu{\`e}re}}, \bibinfo {author} {\bibfnamefont {P.}~\bibnamefont {Menczel}}, \bibinfo {author} {\bibfnamefont {B.}~\bibnamefont {Li}}, \bibinfo {author} {\bibfnamefont {P.}~\bibnamefont {Hopf}}, \bibinfo {author} {\bibfnamefont {G.}~\bibnamefont {Su{\'a}rez}}, \bibinfo {author} {\bibfnamefont {M.}~\bibnamefont {Gali}}, \bibinfo {author} {\bibfnamefont {J.}~\bibnamefont {Lishman}}, \bibinfo {author} {\bibfnamefont {R.}~\bibnamefont {Gadhvi}}, \bibinfo {author} {\bibfnamefont {R.}~\bibnamefont {Agarwal}}, \emph {et~al.},\ }\bibfield  {title} {\bibinfo {title} {Qutip 5: The quantum toolbox in python},\ }\href {https://arxiv.org/abs/2412.04705} {\bibfield  {journal} {\bibinfo  {journal} {arXiv:2412.04705}\ } (\bibinfo {year} {2024})}\BibitemShut {NoStop}%
\bibitem [{\citenamefont {Hertzberg}\ \emph {et~al.}(2021)\citenamefont {Hertzberg}, \citenamefont {Zhang}, \citenamefont {Rosenblatt}, \citenamefont {Magesan}, \citenamefont {Smolin}, \citenamefont {Yau}, \citenamefont {Adiga}, \citenamefont {Sandberg}, \citenamefont {Brink}, \citenamefont {Chow},\ and\ \citenamefont {Orcutt}}]{Hertzberg2021_gauss_spread}%
  \BibitemOpen
  \bibfield  {author} {\bibinfo {author} {\bibfnamefont {J.~B.}\ \bibnamefont {Hertzberg}}, \bibinfo {author} {\bibfnamefont {E.~J.}\ \bibnamefont {Zhang}}, \bibinfo {author} {\bibfnamefont {S.}~\bibnamefont {Rosenblatt}}, \bibinfo {author} {\bibfnamefont {E.}~\bibnamefont {Magesan}}, \bibinfo {author} {\bibfnamefont {J.~A.}\ \bibnamefont {Smolin}}, \bibinfo {author} {\bibfnamefont {J.-B.}\ \bibnamefont {Yau}}, \bibinfo {author} {\bibfnamefont {V.~P.}\ \bibnamefont {Adiga}}, \bibinfo {author} {\bibfnamefont {M.}~\bibnamefont {Sandberg}}, \bibinfo {author} {\bibfnamefont {M.}~\bibnamefont {Brink}}, \bibinfo {author} {\bibfnamefont {J.~M.}\ \bibnamefont {Chow}},\ and\ \bibinfo {author} {\bibfnamefont {J.~S.}\ \bibnamefont {Orcutt}},\ }\bibfield  {title} {\bibinfo {title} {Laser-annealing josephson junctions for yielding scaled-up superconducting quantum processors},\ }\href {https://doi.org/10.1038/s41534-021-00464-5} {\bibfield  {journal} {\bibinfo  {journal} {npj Quantum Information}\ }\textbf {\bibinfo
  {volume} {7}},\ \bibinfo {pages} {129} (\bibinfo {year} {2021})}\BibitemShut {NoStop}%
\bibitem [{\citenamefont {Quantum}(2025)}]{ibm_nairobi}%
  \BibitemOpen
  \bibfield  {author} {\bibinfo {author} {\bibfnamefont {I.}~\bibnamefont {Quantum}},\ }\href {https://quantum.cloud.ibm.com/} {\bibinfo {title} {Ibm quantum processing unit: ibm\_nairobi}} (\bibinfo {year} {2025})\BibitemShut {NoStop}%
\bibitem [{\citenamefont {Hu}\ \emph {et~al.}(2025)\citenamefont {Hu}, \citenamefont {Gomez}, \citenamefont {Chen}, \citenamefont {Trowbridge}, \citenamefont {Goldschmidt}, \citenamefont {Manchester}, \citenamefont {Chong}, \citenamefont {Jaffe},\ and\ \citenamefont {Yelin}}]{hu2025universal}%
  \BibitemOpen
  \bibfield  {author} {\bibinfo {author} {\bibfnamefont {H.-Y.}\ \bibnamefont {Hu}}, \bibinfo {author} {\bibfnamefont {A.~M.}\ \bibnamefont {Gomez}}, \bibinfo {author} {\bibfnamefont {L.}~\bibnamefont {Chen}}, \bibinfo {author} {\bibfnamefont {A.}~\bibnamefont {Trowbridge}}, \bibinfo {author} {\bibfnamefont {A.~J.}\ \bibnamefont {Goldschmidt}}, \bibinfo {author} {\bibfnamefont {Z.}~\bibnamefont {Manchester}}, \bibinfo {author} {\bibfnamefont {F.~T.}\ \bibnamefont {Chong}}, \bibinfo {author} {\bibfnamefont {A.}~\bibnamefont {Jaffe}},\ and\ \bibinfo {author} {\bibfnamefont {S.~F.}\ \bibnamefont {Yelin}},\ }\bibfield  {title} {\bibinfo {title} {Universal dynamics with globally controlled analog quantum simulators},\ }\href {https://doi.org/10.48550/arXiv.0707.1119} {\bibfield  {journal} {\bibinfo  {journal} {arXiv:2508.19075}\ } (\bibinfo {year} {2025})}\BibitemShut {NoStop}%
\bibitem [{\citenamefont {Aiudi}\ \emph {et~al.}()\citenamefont {Aiudi}, \citenamefont {Despres}, \citenamefont {Menta}, \citenamefont {Abedi}, \citenamefont {Menichetti}, \citenamefont {Giovannetti}, \citenamefont {Polini},\ and\ \citenamefont {Caravelli}}]{data}%
  \BibitemOpen
  \bibfield  {author} {\bibinfo {author} {\bibfnamefont {R.}~\bibnamefont {Aiudi}}, \bibinfo {author} {\bibfnamefont {J.}~\bibnamefont {Despres}}, \bibinfo {author} {\bibfnamefont {R.}~\bibnamefont {Menta}}, \bibinfo {author} {\bibfnamefont {A.}~\bibnamefont {Abedi}}, \bibinfo {author} {\bibfnamefont {G.}~\bibnamefont {Menichetti}}, \bibinfo {author} {\bibfnamefont {V.}~\bibnamefont {Giovannetti}}, \bibinfo {author} {\bibfnamefont {M.}~\bibnamefont {Polini}},\ and\ \bibinfo {author} {\bibfnamefont {F.}~\bibnamefont {Caravelli}},\ }\bibfield  {title} {\bibinfo {title} {Data overcoming disorder in superconducting globally driven quantum computing},\ }\href {https://doi.org/10.5281/zenodo.18183082} {\bibinfo  {journal} {zenodo}\ }\BibitemShut {NoStop}%
\bibitem [{\citenamefont {Kingma}\ and\ \citenamefont {Ba}(2014)}]{kingma2014adam}%
  \BibitemOpen
\bibfield  {journal} {  }\bibfield  {author} {\bibinfo {author} {\bibfnamefont {D.~P.}\ \bibnamefont {Kingma}}\ and\ \bibinfo {author} {\bibfnamefont {J.}~\bibnamefont {Ba}},\ }\bibfield  {title} {\bibinfo {title} {Adam: A method for stochastic optimization},\ }\href {https://doi.org/10.48550/arXiv.1412.6980} {\bibfield  {journal} {\bibinfo  {journal} {arXiv:1412.6980}\ } (\bibinfo {year} {2014})}\BibitemShut {NoStop}%
\end{thebibliography}
\end{document}